\documentclass[%
 reprint,
superscriptaddress,
 preprintnumbers,
 nofootinbib,
 amsmath,amssymb,
 aps,
 prd,
]{revtex4-1}

\usepackage{graphicx}
\usepackage{float}
\usepackage{dcolumn}
\usepackage{bm}
\usepackage{color}
\usepackage{amssymb}  
\usepackage{hyperref}
\usepackage{amsmath}
\usepackage{mathtools}
\usepackage{makecell}
\setcellgapes{0.2cm}

\newcommand\numberthis{\addtocounter{equation}{1}\tag{\theequation}}
\allowdisplaybreaks
\begin{document}


\title{Effect of Thomas Rotation on the Lorentz Transformation of Electromagnetic fields}
\author{L.~Malhotra}                  \affiliation{Department of Physics and Astronomy, University of Kentucky, Lexington, Kentucky 40506, USA}
\author{R.~Golub}						\affiliation{Department of Physics, North Carolina State University, Raleigh, North Carolina 27695, USA}
\author{E.~Kraegeloh}			 \affiliation{Department of Physics, University of Michigan, Ann Arbor, Michigan 48109, USA}
\author{N.~Nouri}						\thanks{Present address: Department of Pathology, Yale University School of Medicine, New Haven, CT 06510, USA}\affiliation{Department of Physics and Astronomy, University of Kentucky, Lexington, Kentucky 40506, USA}
\author{B.~Plaster}					\affiliation{Department of Physics and Astronomy, University of Kentucky, Lexington, Kentucky 40506, USA}

\begin{abstract}
	 A relativistic particle undergoing successive boosts which are non collinear will experience a rotation of its coordinate axes with respect to the boosted frame. This rotation of coordinate axes is caused by a relativistic phenomenon called Thomas Rotation. We assess the importance of Thomas rotation in the calculation of physical quantities like electromagnetic fields in the relativistic regime. We calculate the electromagnetic field tensor for general three dimensional successive boosts in the particle's rest frame as well as the laboratory frame. We then compare the electromagnetic field tensors obtained by a direct boost $\vec{\beta} + \delta \vec{\beta}$ and successive boosts $\vec{\beta}$ and $\Delta \vec{\beta}$ and check their consistency with Thomas rotation. This framework might be important to situations such as the calculation of frequency shifts for relativistic spin-1/2 particles undergoing Larmor precession in electromagnetic fields with small field non-uniformities.\\
	
\end{abstract}
\maketitle

\section{Introduction}

As pointed out by Thomas \cite{thomas}, two successive non collinear Lorentz boosts are not equal to a direct boost but to a direct boost followed by a rotation of the coordinate axes. That is,
	\begin{figure}[h]
		\centering
		\includegraphics[scale=0.70]{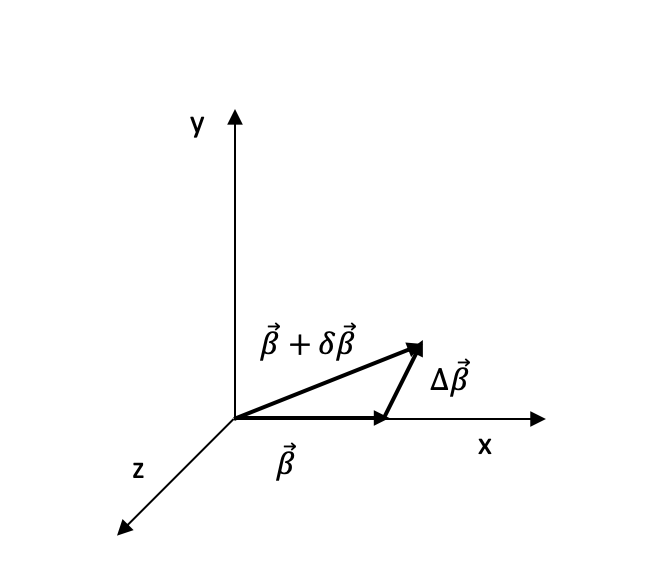} 
		\caption{Schematic of the boosts. $\bm{\vec{\beta}+\delta \vec{\beta}}$: Direct boost, $\bm{\vec{\beta}}$: First successive boost, $\bm{\delta \vec{\beta}}$: Second successive boost in lab frame, $\bm{\Delta \vec{\beta}}$: Second successive boost with respect to the inertial frame with boost $\vec{\beta}$.}
		\label{fig:setup}
	\end{figure}
	\begin{align*}
		A(\vec{\beta}+\delta \vec{\beta}) & \neq A(\delta \vec{\beta}) \cdot A(\vec{\beta})\\ 
		A(\vec{\beta}+\delta \vec{\beta}) &= R_{\mathrm{tom}}(\Delta \vec{\Omega}) \cdot A(\Delta \vec{\beta}) \cdot A(\vec{\beta})  \numberthis \label{eq:1}
	\end{align*} %
 where $(\vec{\beta}+\delta \vec{\beta})$ is the direct boost, $\vec{\beta}$ and $\delta \vec{\beta}$ are two successive boosts in the lab frame, $\Delta \vec{\beta}$ and $\Delta \vec{\Omega}$ are, respectively, the successive boost and the angle of rotation with respect to the frame with boost $\vec{\beta}$. $A(\vec{\beta}+\delta \vec{\beta})$, $A(\vec{\beta} )$  $A(\delta \vec{\beta})$ and $A(\Delta \vec{\beta})$ are the usual boost matrices for the direct boost and the successive boosts respectively, and $R_{\mathrm{tom}}(\Delta \vec{\Omega})$ is the rotation matrix \cite{jackson4, jackson5}.\par
 This rotation of the space coordinates under the application of successive Lorentz boosts is called Thomas rotation. This phenomenon occurs when a relativistic particle is undergoing accelerated motion. Now since we have to show the acceleration, we added an infinitesimal boost vector $\delta \vec{\beta}$ to the original boost $\vec{\beta}$. \par 
In general, for boosts $\vec{\beta}_1$ and $\vec{\beta}_2$ which are parallel to each other or more specifically boosts corresponding to (1+1)–dimensional pure Lorentz transformations, the transformation matrix forms a group which satisfies the equation:
\begin{equation} \label{eq:2}
	A(\vec{\beta}_1) \cdot A(\vec{\beta}_2) = A(\vec{\beta}_{12})  
\end{equation}
where $\beta_{12}$ is the velocity composition of two boosts which is given by the equation:
\begin{equation} \label{eq:3}
	\beta_{12} = \frac{\beta_1 + \beta_2}{1 + \beta_1\cdot \beta_2}
\end{equation}
  But successive boosts which are non collinear, in general, result in Thomas rotation of the space coordinates or in other words, the boosted frames which are accelerating in the sense that their direction is changing will experience Thomas rotation. So the values of the physical quantities obtained by applying just Lorentz transformation are not correct in such cases. \par 
This work is inspired by the ideas discussed in \cite{ungar2,ungar3, ungar4, ungar5, ungar6, ungar7, ungar9} but in a slightly different manner. Ungar et al. defined three inertial reference frames $\Sigma$, $\Sigma'$, and $\Sigma''$ in such a way that their corresponding axes are parallel to each other ($\Sigma$ being the lab frame). It is assumed that the relative velocity of $\Sigma'$ with respect to $\Sigma$ and the relative velocity of $\Sigma''$ with respect to $\Sigma'$ is known beforehand. The relativistic velocity composition law can then be used to calculate the velocity of $\Sigma''$ with respect to $\Sigma$. Usually the velocity of $\Sigma''$ with respect to $\Sigma'$ is not known so the above mentioned approach cannot be used directly. To circumvent this issue, in this paper we present a calculation in which we calculated a matrix $A_T$ \cite{jackson5} from $A(\vec{\beta}+\delta \vec{\beta})$ and $A(-\vec{\beta})$ which contains all the information about relativistic composition of velocities and Thomas rotation.  \par 
To our knowledge, the case of non-collinear boosts and its effects on the electromagnetic field tensor has not been discussed in the literature. The aim of this paper is to see how the electromagnetic field tensor transforms with the Lorentz transformations for general three-dimensional boosts and to show that the field tensor in the direct boosted frame $A(\vec{\beta}+\delta \vec{\beta})$ and successive boosted frames $A(\vec{\beta})$ and $A(\delta \vec{\beta})$ are consistent with Thomas rotation.

\section{Survey of some concepts of the Special Theory of Relativity}
\subsection{Lorentz Transformations}
For two inertial reference frames $\Sigma$ and $\Sigma'$ which have a relative velocity of $\vec{v}$ in such a way that the coordinate axes of $\Sigma$ are parallel to $\Sigma'$ and $\Sigma'$ is moving in the positive \textit{x} direction as seen from $\Sigma$, the position 4-vector of $\Sigma'$ is related to the position 4-vector of $\Sigma$ by the standard Lorentz transformation equations \cite{jackson1}:
\begin{align*}
	x_0' &= \gamma(x_0 - \beta x_1) \\
	x_1' &= \gamma(x_1 - \beta x_0) \\
	x_2' &= x_2 \\
	x_3' &= x_3  \numberthis \label{eq:4}
\end{align*} \par
where \\
$x_0 = ct$, $x_1 = x$, $x_2 = y$, $x_3=z$; \\ \\
 	\hangindent=1.5cm $\vec{\beta} = \frac{\vec{v}}{c}$, $\beta = \mid \vec{\beta} \mid $; \\ \\
 	$\gamma = (1-\beta^2)^{-1/2}$: Lorentz factor  \\ \par
The generalization of Eq. \eqref{eq:4} for the relative velocity of $\Sigma'$ in an arbitrary direction but with the coordinate axes of the two frames still parallel to each other is given by:
\begin{align*}
	x_0' &= \gamma(x_0 - \vec{\beta} \cdot \vec{x}) \\
	\vec{x}' &= \vec{x} + \frac{(\gamma-1)}{\beta^2} \left(\vec{\beta} \cdot \vec{x}\right) \vec{\beta} - \gamma \vec{\beta} x_0 \numberthis \label{eq:5}
\end{align*} \par

\subsection{Addition of Velocities}
Consider two inertial reference frames $\Sigma$ and $\Sigma'$ such that the relative velocity of $\Sigma'$ with respect to $\Sigma$ is $\vec{v}$. A particle is moving in $\Sigma'$ such that its velocity with respect to $\Sigma'$ is $\vec{u}'$. The velocity of the particle with respect to $\Sigma$ is then given by \cite{jackson2}:
	\begin{align*}
		u_{\parallel} &= \frac{u_{\parallel}'+ v}{1+\frac{\vec{v} \cdot \vec{u}'}{c^2}} \\
		\vec{u}_{\perp} &= \frac{\vec{u}'_{\perp}}{\gamma_v (1+\frac{\vec{v} \cdot \vec{u}'}{c^2})} \numberthis \label{eq:6}
	\end{align*} 
where $u_{\parallel}$ and $\vec{u}_{\perp}$ refer to the components of velocity parallel and perpendicular, respectively, to $\vec{v}$. \par
It can be shown that the Lorentz factor of $\vec{v}$, $\vec{u}$, and $\vec{u}'$ are related to each other by 
\begin{equation} \label{eq:7}
	\gamma_u = \gamma_v \gamma_{u'} \left(1+\frac{\vec{v} \cdot \vec{u}'}{c^2}\right)
\end{equation}
More generally, the velocity composition law for two arbitrary velocities can be written as \cite{ungar2,ungar3, ungar4, ungar5, ungar6, ungar7, ungar8, ungar9}:
\begin{equation*} 
\vec{u} \oplus \vec{v} = \frac{\vec{u} + \vec{v}}{1+\frac{\vec{u} \cdot \vec{v}}{c^2}} + \frac{1}{c^2} \left(\frac{\gamma_u}{\gamma_u+1}\right) \frac{\vec{u} \times (\vec{u} \times \vec{v})}{1+\frac{\vec{u} \cdot \vec{v}}{c^2}}
\end{equation*}  \\ 
with
\begin{equation}\label{eq:8}
	 \gamma_{u \oplus v} = \gamma_u \gamma_v  \left(1+\frac{\vec{u} \cdot \vec{v}}{c^2}\right) 
\end{equation}
where symbol $\oplus$ refers to the direct sum of the vector space of the velocity vectors.
\subsection{Matrix Representation and Boost Matrix}
For the rest of the paper, we will be using matrix methods to calculate Lorentz transformations as they are very convenient to use and are more explicit. All of the equations in Eqs. \eqref{eq:4} and \eqref{eq:5} can easily be obtained by using the boost matrices for Lorentz transformations. For example, for a boost along the \textit{x} axis, the boost matrix can be written as \cite{jackson5}:
\begin{equation} \label{eq:9}
	A = \begin{pmatrix}
	\gamma & -\gamma \beta & 0 & 0 \\
	-\gamma \beta & \gamma & 0 & 0 \\
	0 & 0 & 1 & 0 \\
	0 & 0 & 0 & 1 \\ 
	\end{pmatrix}
\end{equation}
Hence,
\begin{align*} 
\begin{pmatrix}
	ct' \\
	x' \\
	y' \\
	z' \\
\end{pmatrix} &= 
\begin{pmatrix}
\gamma & -\gamma \beta & 0 & 0 \\
-\gamma \beta & \gamma & 0 & 0 \\
0 & 0 & 1 & 0 \\
0 & 0 & 0 & 1 \\ 
\end{pmatrix}
\begin{pmatrix}
	ct \\
	x \\
	y \\
	z \\
\end{pmatrix}\\
&= \begin{pmatrix}
	\gamma(ct - \beta x) \\
	\gamma(x - \beta ct) \\
	y \\
	z
\end{pmatrix} \numberthis \label{eq:10}
\end{align*} \par 
For an arbitrary boost, the general form of the boost matrix $A$ takes a form in which the matrix elements can be written as:
\begin{align*} 
	A_{00} &= \gamma \\
	A_{0i} &= A_{i0} = - \gamma \beta_i \\
	A_{ij} &= A_{ji} = \delta_{ij} + (\gamma-1) \frac{\beta_i \beta_j}{\beta^2} \numberthis \label{eq:11}
\end{align*} \par
where $\delta_{ij}$ is the kronecker delta.
The general form of Eq. \eqref{eq:9} can therefore be written as \cite{jackson4}:
\begin{widetext} 
	\[
 \large A = \left(
	\begin{array}{cccc}
	\gamma  & -\gamma \beta_x & -\gamma \beta_y & -\gamma \beta_z \\
	-\gamma \beta_x & 1 + (\gamma-1) \frac{\beta_x^2}{\beta^2} & (\gamma-1) \frac{\beta_x \beta_y}{\beta^2} & (\gamma-1) \frac{\beta_x \beta_z}{\beta^2} \\
	-\gamma \beta_y & (\gamma-1) \frac{\beta_y \beta_x}{\beta^2} & 1 + (\gamma-1) \frac{\beta_y^2}{\beta^2} & (\gamma-1) \frac{\beta_y \beta_z}{\beta^2}  \\
	-\gamma \beta_z & (\gamma-1) \frac{\beta_z \beta_x}{\beta^2} & (\gamma-1) \frac{\beta_z \beta_y}{\beta^2} & 1 + (\gamma-1) \frac{\beta_z^2}{\beta^2} \\
	\end{array} \right)  \numberthis \label{eq:12}
	\]
\end{widetext}

\subsection{Set-up}
To start with, consider two arbitrary boosts $\vec{\beta}$ and $\delta \vec{\beta}$ in three dimensions:
\begin{align*}
	\vec{\beta} &= \beta_x \hat{x} + \beta_y \hat{y} + \beta_z \hat{z} \\
	\delta \vec{\beta} &= \delta \beta_x \hat{x} + \delta \beta_y \hat{y} + \delta \beta_z \hat{z} \numberthis \label{eq:13}
\end{align*}
\begin{figure}[h]
	\centering
	\includegraphics[scale=0.70]{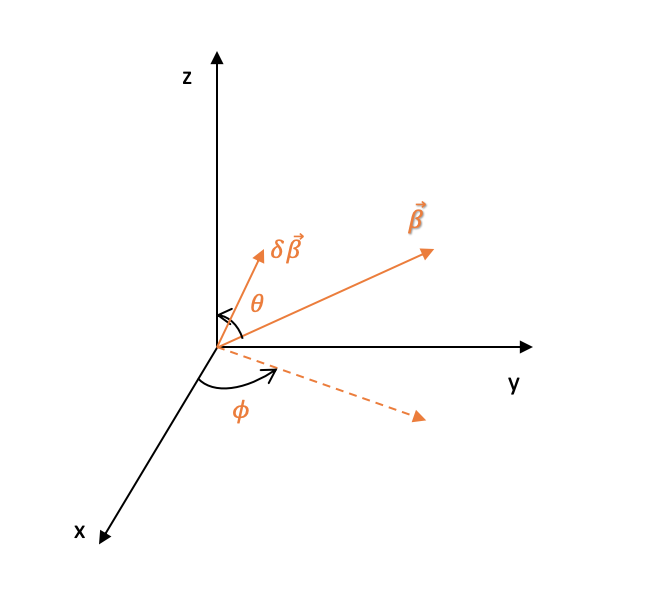} 
	\caption{General boost in three-dimensions. Dotted line represents the projection of $\vec{\beta}$ on the \textit{xy}-plane.}
	\label{fig:boost1}
\end{figure} \par 
In order to calculate the boost matrix for various boosts, we will apply a passive transformation which will rotate our lab frame ($xy$) coordinate axes in such a way that its $x$-axis is aligned with $\vec{\beta}$. This rotated frame will hereafter be called the  longitudinal-transverse ($\ell t$) frame. \par
\begin{figure}[h]
	\centering
	\includegraphics[scale=0.50]{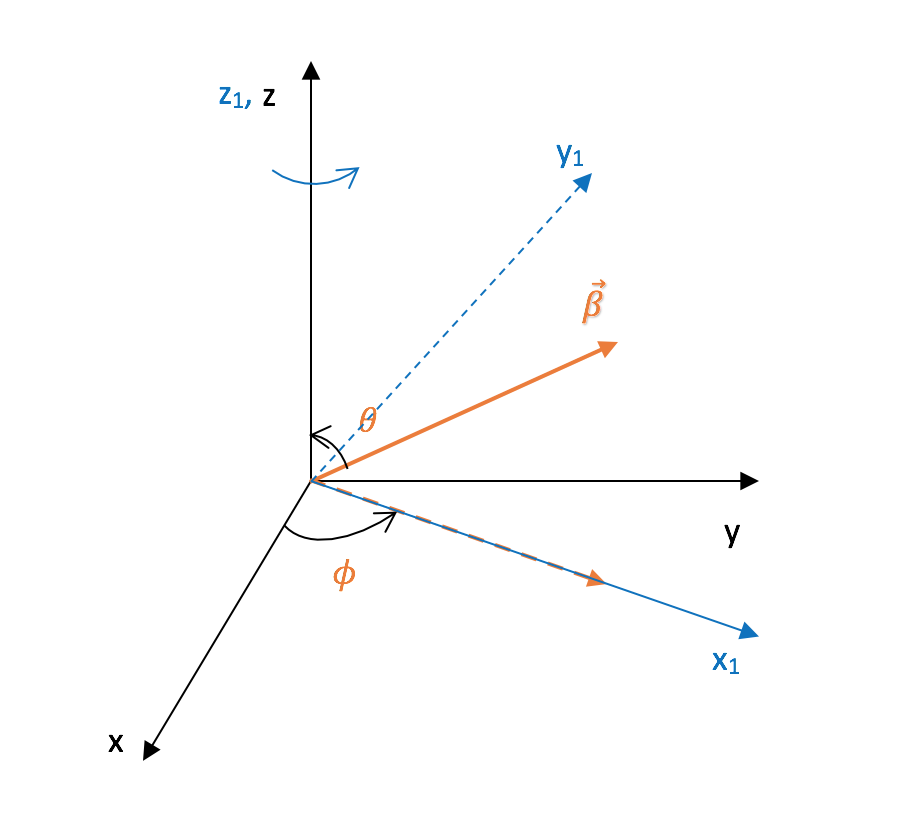} 
	\caption{Rotation about \textit{z} axis by an angle $\phi$. The new \textit{x}, \textit{y} and \textit{z} axes are called the $x_1$, $y_1$ and $z_1$ axes respectively.}
	\label{fig:rotation1}
\end{figure}

 
This whole transformation can be imagined as a product of two rotations: The first rotation is about the \textit{z} axis by an angle $\phi$ which will align the \textit{x} axis along the projection of the boost vector in the \textit{xy} plane (Fig. \ref{fig:rotation1}). 
The rotation matrix associated with this rotation can be written as:
\begin{equation} \label{eq:14}
R_1 = \begin{pmatrix}
1 & 0 & 0 & 0 \\
0 & \cos {\phi} & \sin {\phi}  & 0 \\
0 & - \sin {\phi} &  \cos {\phi} & 0 \\
0 & 0 & 0 & 1 \\ 
\end{pmatrix}
\end{equation} \par
\begin{figure}[h]
	\centering
	\includegraphics[scale=0.50]{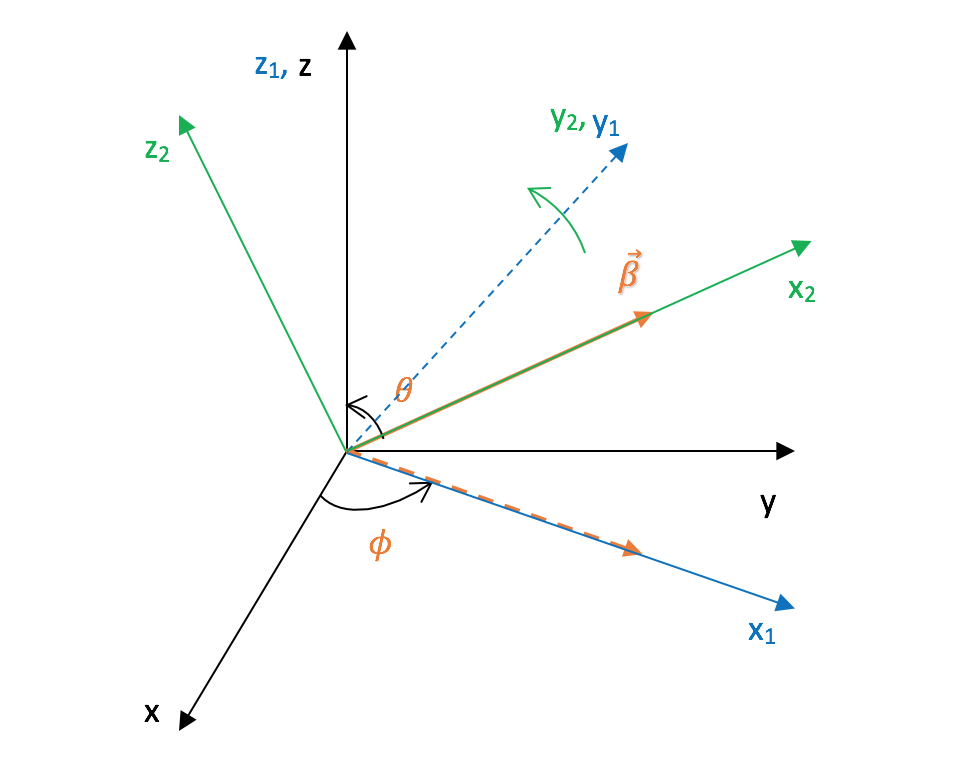} 
	\caption{Second rotation about the $y_1$ axis by an angle $\frac{\pi}{2} - \theta$. $x$, $y$, and $z$ axes in this new frame are called the $x_2$, $y_2$ and $z_2$ axes respectively.}
	\label{fig:rotation2}
\end{figure}

 
The second rotation is about the $y_1$ axis (Fig. \ref{fig:rotation2}) by an angle $\frac{\pi}{2} - \theta$. The effect of this rotation is that it aligns the $x_1$ axis to the boost vector $\vec{\beta}$.
For the second rotation, the rotation matrix can be written as: \\
\begin{equation} \label{eq:15}
R_2 = \begin{pmatrix}
1 & 0 & 0 & 0 \\
0 & \sin {\theta} & 0 & \cos {\theta} \\
0 & 0 & 1 & 0 \\
0 & - \cos {\theta} & 0 & \sin {\theta} \\ 
\end{pmatrix}
\end{equation} \par 
The overall effect of the two rotations can be combined in a single transformation matrix \textit{R}:
\begin{align*}
	R &= R_2 \cdot R_1 \\
		&= \begin{pmatrix}
		1 & 0 & 0 & 0 \\
		0 & \sin {\theta} \cos {\phi} & \sin {\theta} \sin {\phi} & \cos {\theta} \\
		0 & - \sin {\phi} & \cos {\phi} & 0 \\
		0 & - \cos {\theta} \cos {\phi} & - \cos {\theta} \sin {\phi}  & \sin {\theta} \\ 
		\end{pmatrix} \numberthis \label{eq:16}
\end{align*}
It is clear from Fig. \ref{fig:boost1} that if:
\begin{equation*}
	\vec{\beta} = \beta_x \hat{x} + \beta_y \hat{y} + \beta_z \hat{z}
\end{equation*} \\
then 
\begin{align*}
	\cos{\theta} &= \frac{\beta_z}{\lambda_1} \\
	\sin{\theta} &= \frac{\eta_1}{\lambda_1} \\
	\cos{\phi} &= \frac{\beta_x}{\eta_1} \\
	\sin{\phi} &= \frac{\beta_y}{\eta_1} \numberthis \label{eq:17}
\end{align*} 
where the parameters $\lambda_1$ and $\eta_1$ are defined in the appendix. \par 
Using Eq. \eqref{eq:17}, the matrix \textit{R} can be written as:  

\begin{equation} 
	\large R = \left(
	\begin{array}{cccc}
	1 & 0 & 0 & 0 \\[5pt]
	0 & \frac{\beta _x}{\lambda_1} & \frac{\beta _y}{\lambda_1} & \frac{\beta _z}{\lambda_1} \\[5pt]
	0 & -\frac{\beta _y}{\eta_1} & \frac{\beta _x}{\eta_1} & 0 \\[5pt]
	0 & -\frac{\beta _x \beta _z}{\eta_1 \lambda_1} & -\frac{\beta _y \beta _z}{\eta_1 \lambda_1} & \frac{\eta_1}{\lambda_1} \\[5pt]
	\end{array} \right)  \numberthis \label{eq:18}
\end{equation}
As mentioned earlier, this rotation matrix will transform the lab frame coordinates of any 4-vector ($xy$-frame) to its coordinates in the rotating frame also called 
longitudinal-transverse frame ($\ell t$-frame):
\begin{align*}
\vec{\beta}^{\ell t} &= R \cdot \vec{\beta}^{xy} \\
\begin{pmatrix}
0 \\[5pt]
\beta_x^{\ell t} \\[5pt]
\beta_y^{\ell t} \\[5pt]
\beta_z^{\ell t}
\end{pmatrix} &= R \cdot \begin{pmatrix}
0 \\[5pt]
\beta_x^{xy} \\[5pt]
\beta_y^{xy} \\[5pt]
\beta_z^{xy}
\end{pmatrix} \\
 &= \begin{pmatrix}
	0 \\
	\lambda_1 \\
	0 \\
	0 \\
\end{pmatrix}
 \numberthis \label{eq:19}
\end{align*} 
where the superscripts $\ell t$ and $xy$ refer to the components in longitudinal-transverse frame and laboratory frame, respectively, and for the sake of simplicity in notation we assumed:
\begin{equation*}
	\beta_x^{xy} = \beta_x ,\qquad \beta_y^{xy} = \beta_y , \qquad \beta_z^{xy} = \beta_z ;
\end{equation*}\par
Similarly, the infinitesimal boost in the longitudinal-transverse frame is of the form: 
\begin{align*}
	\delta \vec{\beta}^{\ell t} &= R \cdot \delta \vec{\beta}^{xy} \\
	\begin{pmatrix}
	0 \\[5pt]
	\delta \beta_x^{\ell t} \\[5pt]
	\delta \beta_y^{\ell t} \\[5pt]
	\delta \beta_z^{\ell t}
	\end{pmatrix} &= R \cdot \begin{pmatrix}
	0 \\[5pt]
	\delta \beta_x^{xy} \\[5pt]
	\delta \beta_y^{xy} \\[5pt]
	\delta \beta_z^{xy}
	\end{pmatrix} \\
	&= \begin{pmatrix}
	0 \\[5pt]
	\frac{\lambda_2}{\lambda_1} \\[5pt]
	\frac{\lambda_6}{\eta_1} \\[5pt]
	\frac{\lambda_5}{\eta_1 \lambda_1} \\
	\end{pmatrix}
	\numberthis \label{eq:20}
\end{align*} \\ 
where $\lambda_1$, $\lambda_2$, $\lambda_5$, $\lambda_6$ and $\eta_1$ are defined in the Appendix. \par 
To calculate the $\gamma_{(\vec{\beta}+\delta \vec{\beta})}$ in the $\ell t$-frame, we have:
	\begin{equation*}
		(\vec{\beta}+\delta \vec{\beta})^{\ell t} = \left(\lambda_1 + \frac{\lambda_2}{\lambda_1}\right) \hat{x} + \frac{\lambda_6}{\eta_1} \hat{y} + \frac{\lambda_5}{\eta_1 \lambda_1} \hat{z} 
		\end{equation*}
		Keeping the terms linear in $\delta \beta$, we get
		\begin{equation*}
		\mid(\vec{\beta}+\delta \vec{\beta})^{\ell t}\mid^2 \approx \lambda_1^2 + 2 \lambda_2 \\
		\end{equation*}
		Using the above equation we calculate:
		\begin{align*}
			\gamma_{(\vec{\beta}+\delta \vec{\beta})} &= \left(1 - \lambda_1^2 - 2\lambda_2\right)^{-\frac{1}{2}} \\
			& \approx \left(1-\lambda_1^2\right)^{-\frac{1}{2}}\left[1 + \frac{ \lambda_2}{1-\lambda_1^2}\right]
		\end{align*}
		Hence, $\gamma_{(\vec{\beta}+\delta \vec{\beta})} $ can be written as:
		\begin{equation*}
			\gamma_{(\vec{\beta}+\delta \vec{\beta})} \approx \gamma (1 + \gamma^2 \lambda_2)  
		\end{equation*}
	where $\gamma = \frac{1}{\sqrt{1 - \lambda_1^2}}$ is the Lorentz factor.\par 
Using Eq. \eqref{eq:19}, the boost matrix for boost $\vec{\beta}^{\ell t}$ can be calculated:
\begin{equation} \label{eq:21}
A(\vec{\beta})^{\ell t} = 
\begin{pmatrix}
	\gamma  & -\gamma  \lambda_1 & 0 & 0 \\
	-\gamma  \lambda_1 & \gamma  & 0 & 0 \\
	0 & 0 & 1 & 0 \\
	0 & 0 & 0 & 1 \\
\end{pmatrix}
\end{equation}
Similarly, using Eqs. \eqref{eq:19}, \eqref{eq:20}, to the first order in $\delta \beta$, the boost matrix for the direct boost $\vec{\beta} + \delta \vec{\beta}$ in the $\ell t$-frame can be written as: 
\begin{equation} \footnotesize
	A(\vec{\beta} + \delta \vec{\beta})^{\ell t} = 
		\begin{pmatrix}
	\gamma +  \gamma ^3 \lambda_2 & -\frac{\gamma  (\lambda_1^2 + \gamma^2\lambda_2)}{\lambda_1} & -\frac{\gamma \lambda_6}{\eta_1} & -\frac{\gamma  \lambda_5}{\eta_1 \lambda_1}\\[5pt]
	-\frac{\gamma  (\lambda_1^2 + \gamma^2\lambda_2)}{\lambda_1} & \gamma +  \gamma ^3 \lambda_2 & \frac{(\gamma -1) \lambda_6}{\eta_1 \lambda_1} & \frac{(\gamma -1) \lambda_5}{\eta_1 \lambda_1^2}\\[5pt]
	-\frac{\gamma \lambda_6}{\eta_1} & \frac{(\gamma -1) \lambda_6}{\eta_1 \lambda_1} & 1 & 0 \\[5pt]
		-\frac{\gamma  \lambda_5}{\eta_1 \lambda_1} & \frac{(\gamma -1) \lambda_5}{\eta_1 \lambda_1^2} &	0 & 1 
	\end{pmatrix} \numberthis \label{eq:22}
\end{equation}

\section{Transformations of the Electromagnetic Field Tensor}
The main idea of this paper is to see how the electromagnetic fields transform relativistically when there is an accelerated motion. It can be further divided into transformations in the longitudinal-transverse and lab frame.
\subsection{Longitudinal-Transverse \boldmath{$\ell t$}-Frame}
To see the effects on electromagnetic fields, we first need to bring the electromagnetic field tensor to the rotating $\ell t$-frame so that all the boosts and electromagnetic fields are in the same frame to start with. The electromagnetic field tensor $F^{\mu \nu}$ in the lab frame is given by \cite{jackson6}:
\begin{equation} \label{eq:23}
	F^{\mu \nu} = \begin{pmatrix}
	0 & -E_x & -E_y & -E_z \\
	E_x & 0 & -B_z & B_y \\
	E_y & B_z & 0 & -B_x \\
	E_z & -B_y & B_x & 0
	\end{pmatrix}
\end{equation} 
For the rest of the paper we will write $F^{\mu \nu} = F$. \par 
To get the field tensor in the $\ell t$-frame, we can apply the rotation matrix \textit{R} on $F^{\mu \nu}$:
\begin{equation} \label{eq:24}
	F^{\ell t} = R \cdot F \cdot R^T
\end{equation}\\
where the superscript $T$ refers to matrix transpose. After plugging in the values of \textit{R} and \textit{F} from Eqs. \eqref{eq:18} and \eqref{eq:23}, we can write $F^{\ell t}$ as: \\
\begin{widetext}
	\[
	F^{\ell t} = \begin{pmatrix}
	0 & \frac{-\beta _x E_x-\beta _y E_y-\beta _z E_z}{\lambda_1} & \frac{\beta _y E_x-\beta _x E_y}{\eta_1} & \frac{-E_z \beta _x^2+\beta _z E_x \beta _x+\beta _y \left(\beta _z E_y-\beta _y E_z\right)}{\eta_1 \lambda_1} \\
	\frac{\beta _x E_x+\beta _y E_y+\beta _z E_z}{\lambda_1} & 0 & \frac{\left(B_x \beta _x+B_y \beta _y\right) \beta _z-B_z \eta_1^2}{\eta_1 \lambda_1} & \frac{B_y \beta _x-B_x \beta _y}{\eta_1} \\
	\frac{\beta _x E_y-\beta _y E_x}{\eta_1} & \frac{B_z \eta_1^2-\left(B_x \beta _x+B_y \beta _y\right) \beta _z}{\eta_1 \lambda_1} & 0 & \frac{-B_x \beta _x-B_y \beta _y-B_z \beta _z}{\lambda_1} \\
	\frac{E_z \beta _x^2-\beta _x \beta _z E_x+\beta _y \left(\beta _y E_z-\beta _z E_y\right)}{\eta_1 \lambda_1} & \frac{B_x \beta _y-B_y \beta _x}{\eta_1} & \frac{B_x \beta _x+B_y \beta _y+B_z \beta _z}{\lambda_1} & 0 \\
	\end{pmatrix} 
	\]
\end{widetext} \par 
To the electromagnetic field tensor obtained in Eq. \eqref{eq:24}, we will apply boost matrix for the first successive boost $\vec{\beta}^{\ell t}$ and the direct boost $(\vec{\beta} + \delta \vec{\beta})^{\ell t}$ using the well-known equation \cite{jackson7}:
\begin{equation} \label{eq:25}
	F' = A \cdot F \cdot A^T
\end{equation}
where $F'$ and $F$ are the electromagnetic field tensors in the boosted frame and the lab frame (or any inertial frame) respectively and \textit{A} is the boost matrix. \par 
 For boost $\vec{\beta}^{\ell t}$, the transformation of $F^{lt}$ can be calculated using Eqs. \eqref{eq:21}, \eqref{eq:24} and \eqref{eq:25}:
	\begin{equation*} 
		(F')^{\ell t} = A(\vec{\beta}^{\ell t}) \cdot F^{\ell t} \cdot (A(\vec{\beta}^{\ell t}))^T
 	\end{equation*}
 	The following table has the elements of $(F')^{\ell t}$ after matrix multiplication:
	 \begin{widetext}
	 	\begin{center}
	 	\begin{table}[h]
	 		\makegapedcells \begin{tabular}{|c|c|c|}
	 			\hline 
	 			$(F'^{10})^{\ell t}$ & $(E'_x)^{\ell t}$ & $\frac{\beta_x E_x + \beta_y E_y + \beta_z E_z}{\lambda_1}$ \\
	 			\hline
	 			$(F'^{20})^{\ell t}$ & $(E'_y)^{\ell t}$ & $\frac{\gamma  \left(-B_z \eta_1^2+B_x \beta _x \beta _z+B_y \beta _y \beta _z-E_x \beta _y+\beta _x E_y\right)}{\eta_1}$  \\
	 			\hline
	 			 $(F'^{30})^{\ell t}$ & $(E'_z)^{\ell t}$ & $\frac{\gamma  \left(B_y \beta _x \lambda_1^2-B_x \beta _y \lambda_1^2+\beta _x^2 E_z-\beta _x E_x \beta _z-\beta _y E_y \beta _z+\beta _y^2 E_z\right)}{\eta_1 \lambda_1}$ \\
	 			\hline 
	 			 $(F'^{32})^{\ell t}$ & $(B'_x)^{\ell t}$ & $\frac{B_x \beta _x+B_y \beta _y+B_z \beta _z}{\lambda_1}$ \\
	 			\hline
	 			$(F'^{13})^{\ell t}$ & $(B'_y)^{\ell t}$ & $\frac{\gamma  \left(B_y \beta _x-B_x \beta _y+\beta _x^2 E_z-\beta _x E_x \beta _z-\beta _y E_y \beta _z+\beta _y^2 E_z\right)}{\eta_1}$ \\
	 			\hline
	 			$(F'^{21})^{\ell t}$ & $(B'_z)^{\ell t}$ & $\frac{\gamma  \left(B_z \eta_1^2-B_x \beta _x \beta _z-B_y \beta _y \beta _z-\beta _x^3 E_y+\beta _x^2 E_x \beta _y-\beta _x \beta _y^2 E_y+E_x \beta _y^3-\beta _x E_y \beta _z^2+E_x \beta _y \beta _z^2\right)}{\eta_1 \lambda_1}$ \\
	 			\hline
	 		\end{tabular}
	 		\caption{\label{tab:table1}Expressions for the indicated components of the electromagnetic field tensor after being transformed by the first boost $\vec{\beta}^{\ell t}$ in the longitudinal-transverse frame.}
	 	\end{table}
 	\end{center}
	 \end{widetext} 
 Electromagnetic fields in Table \ref{tab:table1} are consistent with the standard field transformation equations \cite{jackson7, ungar5}. 
 \begin{align*}
 	\vec{E}' &= \gamma \left(\vec{E} + \vec{\beta} \times \vec{B} \right) - \frac{\gamma^2}{\gamma + 1}\, \vec{\beta} (\vec{\beta} \cdot \vec{E})  \\
 	\vec{B}' &= \gamma \left(\vec{B} - \vec{\beta} \times \vec{E} \right) - \frac{\gamma^2}{\gamma + 1}\, \vec{\beta} (\vec{\beta} \cdot \vec{B}) \numberthis \label{eq:26}
 \end{align*} 
 Similarly, for the direct boost $(\vec{\beta} + \delta \vec{\beta})^{\ell t}$, the electromagnetic field tensor transformation is given by:
 \begin{equation*}
 	(F'')^{\ell t} = A((\vec{\beta}+\delta \vec{\beta})^{\ell t}) \cdot F^{\ell t} \cdot A((\vec{\beta}+\delta \vec{\beta})^{\ell t})^T
 \end{equation*}
 After simplification and keeping terms to linear order in $\delta \beta $, $(F'')^{\ell t}$ can be calculated and the detailed expressions of its elements are provided in the Appendix (Section A.1). \par 
Because of the way we set up the problem, the electromagnetic field tensor described by the direct boost $(\vec{\beta} + \delta \vec{\beta})^{\ell t}$ already consists of rotations. To get the electromagnetic fields which do not have any rotations (pure Lorentz boost) \cite{jackson5}, we will use the successive boosts. As mentioned earlier in the Introduction, we will calculate a matrix $A_T$:
\begin{equation} \label{eq:27}
	A_T = A(\vec{\beta} + \delta \vec{\beta}) \cdot A(-\vec{\beta})
\end{equation}
For the $\ell t$-frame, $A_T$ looks like: \\
	\begin{equation} \label{eq:28}
		A_T^{\ell t} = \begin{pmatrix}
		1 & \frac{-\gamma ^2 \lambda_2}{\lambda_1} & -\frac{\gamma  \lambda_6}{\eta_1} & -\frac{\gamma  \lambda_5}{\eta_1 \lambda_1}\\[5pt]
		\frac{-\gamma ^2 \lambda_2}{\lambda_1} & 1 & \frac{(\gamma -1) \lambda_6}{\eta_1 \lambda_1} & \frac{(\gamma -1) \lambda_5}{\eta_1 \lambda_1^2} \\[5pt]
		-\frac{\gamma  \lambda_6}{\eta_1} & -\frac{(\gamma - 1) \lambda_6}{\eta_1 \lambda_1} & 1 & 0 \\[5pt]
		-\frac{\gamma  \lambda_5}{\eta_1 \lambda_1} & -\frac{(\gamma -1) \lambda_5}{\eta_1 \lambda_1^2} & 0 & 1
		\end{pmatrix}
	\end{equation}   \par
The matrix $A_T$ contains all the information regarding relativistic composition of velocities and Thomas rotation which can be seen if we write $A_T$ as \cite{jackson5}:
\begin{align*} 
	A_T^{\ell t} &= A(\Delta \vec{\beta})^{\ell t} \cdot R_{\mathrm{tom}}(\Delta \vec{\Omega})^{\ell t}  \\
	& = \left(I - \Delta \vec{\beta}^{\ell t} \cdot \vec{K}\right) \cdot \left(I - \Delta \vec{\Omega}^{\ell t} \cdot \vec{S}\right) \numberthis \label{eq:29}
\end{align*}
where \\
	$\Delta \vec{\beta}^{\ell t}$ = successive boost with respect to frame with boost $\vec{\beta}$; \\
	$\Delta \vec{\Omega}^{\ell t} = \left[ \left(\frac{\gamma -1}{\beta^2}\right) \vec{\beta}^{\ell t} \times \delta \vec{\beta}^{\ell t} \right]$ is the angle of rotation associated with Thomas rotation. \par
	It can be easily shown that if the boosts and rotations are infinitesimal then:
	\begin{center}
		$A(\Delta \vec{\beta}) \cdot R_{\mathrm{tom}}(\Delta \vec{\Omega})  = R_{\mathrm{tom}}(\Delta \vec{\Omega}) \cdot A(\Delta \vec{\beta})$ 
	\end{center}
Matrices $\vec{K}$ and $\vec{S}$ are the generators of Lorentz boosts and rotations respectively:
\begin{align*}
	K_1 &= \begin{pmatrix}
	0 & 1 & 0 & 0 \\
	1 & 0 & 0 & 0 \\
	0 & 0 & 0 & 0 \\
	0 & 0 & 0 & 0 \\
	\end{pmatrix} , \qquad
	K_2 = \begin{pmatrix}
	0 & 0 & 1 & 0 \\
	0 & 0 & 0 & 0 \\
	1 & 0 & 0 & 0 \\
	0 & 0 & 0 & 0 \\
	\end{pmatrix} , \\
	K_3 &= \begin{pmatrix}
	0 & 0 & 0 & 1 \\
	0 & 0 & 0 & 0 \\
	0 & 0 & 0 & 0 \\
	1 & 0 & 0 & 0 \\
	\end{pmatrix} , \qquad
	S_1 =  \begin{pmatrix}
	0 & 0 & 0 & 0 \\
	0 & 0 & 0 & 0 \\
	0 & 0 & 0 & -1 \\
	0 & 0 & 1 & 0 \\
	\end{pmatrix} , \\
	S_2 &=  \begin{pmatrix}
	0 & 0 & 0 & 0 \\
	0 & 0 & 0 & 1 \\
	0 & 0 & 0 & 0 \\
	0 & -1 & 0 & 0 \\
	\end{pmatrix} , \qquad
	S_3 =  \begin{pmatrix}
	0 & 0 & 0 & 0 \\
	0 & 0 & -1 & 0 \\
	0 & 1 & 0 & 0 \\
	0 & 0 & 0 & 0 \\
	\end{pmatrix} 
\end{align*} \par 
Extracting the matrix form of $A(\Delta \vec{\beta}^{\ell t})$ and $R_{\mathrm{tom}}(\Delta \vec{\Omega}^{\ell t})$ from $A_T$, we get:
\begin{equation*}
	A(\Delta \vec{\beta})^{\ell t} = \begin{pmatrix}
	1 & \frac{-\gamma ^2 \lambda_2}{\lambda_1} & -\frac{\gamma  \lambda_6}{\eta_1} & -\frac{\gamma  \lambda_5}{\eta_1 \lambda_1}\\[5pt]
	\frac{-\gamma ^2 \lambda_2}{\lambda_1} & 1 & 0 & 0 \\[5pt]
	-\frac{\gamma  \lambda_6}{\eta_1} & 0 & 1 & 0 \\[5pt]
	-\frac{\gamma  \lambda_5}{\eta_1 \lambda_1} & 0 & 0 & 1
	\end{pmatrix} 
	\end{equation*}
	\begin{equation} \label{eq:30}
	R_{\mathrm{tom}}(\Delta \vec{\Omega})^{\ell t}  = \begin{pmatrix}
		1 & 0 & 0 & 0 \\[5pt]
		0 & 1 & \frac{(\gamma -1) \lambda_6}{\eta_1 \lambda_1} & \frac{(\gamma -1) \lambda_5}{\eta_1 \lambda_1^2} \\[5pt]
		0 & -\frac{(\gamma - 1) \lambda_6}{\eta_1 \lambda_1} & 1 & 0 \\[5pt]
		0 & -\frac{(\gamma -1) \lambda_5}{\eta_1 \lambda_1^2} & 0 & 1
	\end{pmatrix} 
\end{equation} \par 
In order to find the electromagnetic fields due to pure Lorentz boosts, we calculate the electromagnetic field tensor due to the successive boosts $\vec{\beta}^{\ell t}$ and $\Delta \vec{\beta}^{\ell t}$:
\begin{align*} 
	(F''')^{\ell t} &= A(\Delta \vec{\beta})^{\ell t}\cdot A(\vec{\beta})^{\ell t} \cdot F^{\ell t}\cdot (A(\vec{\beta})^{\ell t})^T \cdot (A(\Delta \vec{\beta})^{\ell t})^T \\
	&= A(\Delta \vec{\beta})^{\ell t}\cdot (F')^{\ell t}\cdot (A(\Delta \vec{\beta})^{\ell t})^T\numberthis \label{eq:31}
\end{align*}
After simplification and keeping the terms which are linear in $\delta \beta$, we get the matrix $(F''')^{\ell t}$ whose elements are provided in the Appendix (Section A.2). \par 
It should be noted that since $(F''')^{\ell t}$ and $(F'')^{\ell t}$ are different from each other by just a rotation, so $(F''')^{\ell t}$ can be obtained by operating an inverse Thomas rotation on $(F'')^{\ell t}$. In fact, we used this as a check for verifying if the expressions of electromagnetic fields calculated using Eq. \eqref{eq:32} are correct.
\begin{equation} \label{eq:32}
	(F''')^{\ell t} = R_{\mathrm{tom}}(-\Delta \vec{\Omega})^{\ell t} \cdot (F'')^{\ell t} \cdot (R_{\mathrm{tom}}(-\Delta \vec{\Omega})^{\ell t})^T
\end{equation}

\subsection{Laboratory \textbf{\textit{xy}}-Frame}
After getting the expressions of electromagnetic fields in the $\ell t$-frame obtained by different boosts, we now calculate the electromagnetic fields by the same boosts with respect to the lab frame. The overall approach stays the same but all the boost matrices are  needed to be transformed in the $xy$-frame before being used to calculate the electromagnetic field tensor. Another way of calculating the electromagnetic field tensor is to directly transform the field tensors obtained in $\ell t$-frame. \par 
In order to calculate the electromagnetic field tensor for various boosts in the lab $xy$-frame, we will just use the field tensor $F$ as defined in Eq. \eqref{eq:23}. Since $R$ is the rotation matrix for passive coordinate transformations \eqref{eq:18}, we have:
\begin{equation*}
	R \cdot R^T = R^T \cdot R = I \numberthis \label{eq:33}
\end{equation*}
therefore we can write the electromagnetic tensors and boost matrices in the lab $xy$-frame as:
\begin{align*} 
	F^{xy} &= R^T \cdot F^{\ell t} \cdot R \\
	A^{xy} &= R^T \cdot A^{\ell t} \cdot R \numberthis \label{eq:34}
\end{align*}
After matrix multiplication, $A(\vec{\beta})^{xy}$ can be written as:
\begin{equation*}\label{eq:35}
	A({\vec{\beta}})^{xy} =
	\begin{pmatrix}
	\gamma  & -\gamma  \beta _x & -\gamma  \beta _y & -\gamma  \beta _z \\[5pt]
	-\gamma  \beta _x & \frac{\gamma  \beta _x^2+\beta _y^2+\beta _z^2}{\lambda_1^2} & \frac{(\gamma -1) \beta _x \beta _y}{\lambda_1^2} & \frac{(\gamma -1) \beta _x \beta _z}{\lambda_1^2} \\[5pt]
	-\gamma  \beta _y & \frac{(\gamma -1) \beta _x \beta _y}{\lambda_1^2} & \frac{\beta _x^2+\gamma  \beta _y^2+\beta _z^2}{\lambda_1^2} & \frac{(\gamma -1) \beta _y \beta _z}{\lambda_1^2} \\[5pt]
	-\gamma  \beta _z & \frac{(\gamma -1) \beta _x \beta _z}{\lambda_1^2} & \frac{(\gamma -1) \beta _y \beta _z}{\lambda_1^2} & \frac{\beta _x^2+\beta _y^2+\gamma  \beta _z^2}{\lambda_1^2} \\
	\end{pmatrix}
\end{equation*}
which is in agreement with Eq. \eqref{eq:12} if we substitute in\\
\begin{equation*}
\vec{\beta} = \vec{\beta}^{xy} =  \beta_{x} \hat{x} + \beta_{y} \hat{y} + \beta_{z} \hat{z} 
\end{equation*} \par 
Using Eqs. \eqref{eq:25} and \eqref{eq:34} we can calculate the electromagnetic field tensor $(F')^{xy}$ which corresponds to the boost $\vec{\beta}^{xy}$: \\
\begin{widetext}
	\begin{center}
	\begin{table}[h]
			\makegapedcells\begin{tabular}{|c|c|c|}
				\hline 
				$(F'^{10})^{xy}$ & $(E'_x)^{xy}$ & $\gamma  E_x + \gamma \left( B_z \beta _y- B_y \beta _z\right) -\frac{\gamma ^2 \beta _x \left(\beta _x E_x+\beta _y E_y+\beta _z E_z\right)}{\gamma +1}$ \\
				\hline
				$(F'^{20})^{xy}$ & $(E'_y)^{xy}$ & $\gamma  E_y + \gamma \left( B_x \beta _z- B_z \beta _x\right) -\frac{\gamma ^2 \beta _y \left(\beta _x E_x+\beta _y E_y+\beta _z E_z\right)}{\gamma +1}$ \\
				\hline
				$(F'^{30})^{xy}$ & $(E'_z)^{xy}$ & $\gamma  E_z + \gamma  \left(B_y \beta _x-B_x \beta _y\right)-\frac{\gamma ^2 \beta _z \left(\beta _x E_x+\beta _y E_y+\beta _z E_z\right)}{\gamma +1}$ \\
				\hline 
				$(F'^{32}){xy}$ & $(B'_x)^{xy}$ & $\gamma  B_x +\gamma  \left(E_y \beta _z-\beta _y E_z\right)-\frac{\gamma^2 \beta _x \left(B_x \beta _x+B_y \beta _y+B_z \beta _z\right)}{\gamma +1}$ \\
				\hline
				$(F'^{13})^{xy}$ & $(B'_y)^{xy}$ & $\gamma  B_y +\gamma  \left(E_z \beta _x-\beta_z E_x\right)-\frac{\gamma^2 \beta _x \left(B_x \beta _x+B_y \beta _y+B_z \beta _z\right)}{\gamma +1}$ \\
				\hline
				$(F'^{21})^{xy}$ & $(B'_z)^{xy}$ & $\gamma  B_z +\gamma  \left(E_x \beta _y-\beta _x E_y\right)-\frac{\gamma^2 \beta _x \left(B_x \beta _x+B_y \beta _y+B_z \beta _z\right)}{\gamma +1}$ \\
				\hline
			\end{tabular}
		\caption{\label{tab:table2}Expressions for the indicated components of the electromagnetic field tensor after being transformed by the first boost $\vec{\beta}^{xy}$ in the laboratory frame.}
	\end{table}
\end{center}
\end{widetext} \par
Again, the components of the electromagnetic field tensor in Table \ref{tab:table2} can be verified from the standard field transformations as shown in Eq. \eqref{eq:26}. \par  
Similarly, for the direct boost $(\vec{\beta} + \delta \vec{\beta})^{xy}$, we can use Eq. \eqref{eq:34} to calculate the boost matrix.
The detailed expression of $A(\vec{\beta} + \delta \vec{\beta})^{xy}$ is too long to write here but it shares the same features as \cite{jackson5} which can be seen if we let $\delta \beta_z = \beta_y = \beta_z = 0$:
\begin{widetext}
	\begin{equation} \label{eq:36}
	A(\vec{\beta}+\delta \vec{\beta})^{xy} = \begin{pmatrix}
	\gamma + \gamma^3 \beta_x \delta \beta_x & -(\gamma \beta_x + \gamma^3 \delta \beta_x) & -\gamma \delta \beta_y & 0 \\
	-(\gamma \beta_x + \gamma^3 \delta \beta_x) & \gamma + \gamma^3 \beta_x \delta \beta_x & \left(\frac{\gamma -1}{\beta_x^2}\right) \beta_x \delta \beta_y & 0 \\
	- \gamma \delta \beta_y & \left(\frac{\gamma -1}{\beta_x^2}\right) \beta_x \delta \beta_y & 0 & 1 \\
	0 & 0 & 0 & 1
	\end{pmatrix}
	\end{equation}
\end{widetext}
The above matrix is identical in form to the one shown in \cite{jackson5}. After calculating the boost matrix Eq. \eqref{eq:36}, we can again use Eq. \eqref{eq:25} to calculate the electromagnetic field tensor in the direct boosted frame with respect to the laboratory frame whose detailed expressions are provided in the Appendix (Section B.1).  \par 
In order to calculate electromagnetic fields in the inertial frames which are boosted upon by pure Lorentz boosts (no rotation), we use successive boosts $\vec{\beta}$ and $\Delta \vec{\beta}$. For that we have to calculate the expression of $A_T^{xy}$ first as done in Eq. \eqref{eq:27} which is:
\begin{widetext}
\begin{equation} \label{eq:37}
	A_T^{xy} = \\ 
	\begin{pmatrix}
	1 & -\frac{1}{\lambda_1^2}(\gamma  \lambda_3+\gamma ^2 \beta _x \lambda_2) & 	-\frac{1}{\lambda_1^2}(\gamma  \lambda_4+\gamma ^2 \beta _y \lambda_2) & -\frac{1}{\lambda_1^2}(\gamma  \lambda_5+\gamma ^2 \beta _z \lambda_2)\\[5pt]
	-\frac{1}{\lambda_1^2}(\gamma  \lambda_3+\gamma ^2 \beta _x \lambda_2) & 1 & \frac{(\gamma -1) \lambda_6}{\lambda_1^2} & -\frac{(\gamma -1) \lambda_8}{\lambda_1^2} \\[5pt]   
	-\frac{1}{\lambda_1^2}(\gamma  \lambda_4+\gamma ^2 \beta _y \lambda_2)& -\frac{(\gamma -1) \lambda_6}{\lambda_1^2} & 1 & \frac{(\gamma -1) \lambda_7}{\lambda_1^2}\\[5pt]
	-\frac{1}{\lambda_1^2}(\gamma  \lambda_5+\gamma ^2 \beta _z \lambda_2)& \frac{(\gamma -1) \lambda_8}{\lambda_1^2} & -\frac{(\gamma -1) \lambda_7}{\lambda_1^2} & 1
	\end{pmatrix}
\end{equation} 
\end{widetext}\par 
As we know from Eq. \eqref{eq:29}, $A(\Delta \vec{\beta})^{xy}$ and $R_{\mathrm{tom}}(\Delta \vec{\Omega})^{xy}$ can be extracted from $A_T^{xy}$ which can be written as:
\begin{widetext}
	\begin{equation*}
	A(\Delta \vec{\beta})^{xy} =  \begin{pmatrix}
	1 & -\frac{1}{\lambda_1^2}(\gamma  \lambda_3+\gamma ^2 \beta _x \lambda_2) & 	-\frac{1}{\lambda_1^2}(\gamma  \lambda_4+\gamma ^2 \beta _y \lambda_2) & -\frac{1}{\lambda_1^2}(\gamma  \lambda_5+\gamma ^2 \beta _z \lambda_2)\\[5pt]
	-\frac{1}{\lambda_1^2}(\gamma  \lambda_3+\gamma ^2 \beta _x \lambda_2) & 1 & 0 & 0 \\[5pt]   
	-\frac{1}{\lambda_1^2}(\gamma  \lambda_4+\gamma ^2 \beta _y \lambda_2)& 0 & 1 & 0\\[5pt]
	-\frac{1}{\lambda_1^2}(\gamma  \lambda_5+\gamma ^2 \beta _z \lambda_2)& 0 & 0 & 1
	\end{pmatrix}
	\end{equation*}
	\begin{equation}\label{eq:38}
	 R_{\mathrm{tom}}(\Delta \vec{\Omega})^{xy} =  \begin{pmatrix}
	 1 & 0 & 0 & 0\\[5pt]
	 0 & 1 & \frac{(\gamma -1) \lambda_6}{\lambda_1^2} & -\frac{(\gamma -1) \lambda_8}{\lambda_1^2} \\[5pt]   
	 0 & -\frac{(\gamma -1) \lambda_6}{\lambda_1^2} & 1 & \frac{(\gamma -1) \lambda_7}{\lambda_1^2}\\[5pt]
	 0 & \frac{(\gamma -1) \lambda_8}{\lambda_1^2} & -\frac{(\gamma -1) \lambda_7}{\lambda_1^2} & 1
	 \end{pmatrix}
	\end{equation}
\end{widetext}
where $\lambda_i$, ($i=1,2,\cdots, 8$) and $\eta_1$ are defined in the appendix.\par
Using Eq. \eqref{eq:38} we can now calculate electromagnetic fields due to pure Lorentz boosts whose detailed expressions are provided in the Appendix (Section B.2).
\begin{widetext}
\begin{align*} 
 (F''')^{xy}  &= A(\Delta \vec{\beta})^{xy}\cdot A(\vec{\beta})^{xy}  \cdot F^{xy}\cdot (A(\vec{\beta})^{xy})^T \cdot (A(\Delta \vec{\beta})^{xy})^T  \\
&= A(\Delta \vec{\beta})^{xy}\cdot (F')^{xy}\cdot (A(\Delta \vec{\beta})^{xy})^T\numberthis \label{eq:39}
\end{align*}
\end{widetext}
\section{Validation of Results}
All the framework that we have constructed can be verified by two ways:
\begin{enumerate}
	\item Verifying the form of boost matrices and electromagnetic field tensor for some special cases as discussed in \cite{jackson5, jackson6}. 
	\item Applying this whole formalism on a 4-vector like position.
\end{enumerate}  \par 
For the first approach, in order to see the identical nature of results we will assume the special case of  $\beta_y = \beta_z = \delta \beta_z = 0$. Applying this assumption on Eqs. \eqref{eq:21} and \eqref{eq:22} will give us:
\begin{widetext}
\begin{equation*} 
A(\vec{\beta})^{\ell t} = 
\begin{pmatrix}
\gamma  & -\gamma  \beta_x  & 0 & 0 \\
-\gamma  \beta_x & \gamma  & 0 & 0 \\
0 & 0 & 1 & 0 \\
0 & 0 & 0 & 1 \\
\end{pmatrix}
\end{equation*}
	\begin{equation*} 
	A(\vec{\beta}+\delta \vec{\beta})^{\ell t} = \begin{pmatrix}
	\gamma + \gamma^3 \beta_x \delta \beta_x & -(\gamma \beta_x + \gamma^3 \delta \beta_x) & -\gamma \delta \beta_y & 0 \\
	-(\gamma \beta_x + \gamma^3 \delta \beta_x) & \gamma + \gamma^3 \beta_x \delta \beta_x & \left(\frac{\gamma -1}{\beta_x}\right) \delta \beta_y & 0 \\
	- \gamma \delta \beta_y & \left(\frac{\gamma -1}{\beta_x}\right) \delta \beta_y & 0 & 1 \\
	0 & 0 & 0 & 1
	\end{pmatrix} \numberthis \label{eq:40}
	\end{equation*} 
	\end{widetext}
Similarly, $A_T^{\ell t}$ can be reduced to a familiar result \cite{jackson5}:
\begin{equation}\label{eq:41}
	A_T^{\ell t} = \begin{pmatrix}
	1 & -\gamma ^2 \delta \beta _x  & -\gamma \delta \beta _y & 0 \\
	-\gamma ^2 \delta \beta _x  & 1 & \frac{(\gamma -1)\delta \beta _y}{\beta _x} & 0\\
	-\gamma \delta \beta _y & -\frac{(\gamma -1)\delta \beta _y}{\beta _x} & 1 & 0 \\
	0 & 0 & 0 & 1 \\
	\end{pmatrix}
\end{equation} \par 
In the lab $xy$-frame, we get the exact same results as Eqs. \eqref{eq:40} and \eqref{eq:41} for the above mentioned special case. This makes perfect sense since letting $\beta_y = \beta_z = \delta \beta_z = 0$ would just make the original passive coordinate transformations redundant and both the $\ell t$- and $xy$- frames will be identical. \par 
To see if the matrix for Thomas rotation $R_{\mathrm{tom}}(\Delta \vec{\Omega})$ is correct we can directly calculate it from its definition:
\begin{equation} \label{eq:42}
	R_{\mathrm{tom}}(\Delta \vec{\Omega}) = \left(I - \Delta \vec{\Omega} \cdot \vec{S}\right)
\end{equation}
where \\
\indent \indent \indent  $\Delta \vec{\Omega} = \left[ \left(\frac{\gamma -1}{\beta^2}\right) \vec{\beta} \times \delta \vec{\beta} \right]$ \\
The Thomas rotation matrix calculated from the Eq. \eqref{eq:42} using the corresponding representations of the boost vectors in ${\ell t}$/$xy$ -frames matches with Eqs. \eqref{eq:30} and \eqref{eq:38}. \par 
For the verification of Electromagnetic Field Tensors, we can calculate them in different ways. As an example, we calculated $(F''')^{xy}$ using:
\begin{widetext}
\begin{align*}
	(F''')^{xy}  &= A(\Delta \vec{\beta})^{xy}\cdot A(\vec{\beta})^{xy}  \cdot F^{xy}\cdot (A(\vec{\beta})^{xy})^T \cdot (A(\Delta \vec{\beta})^{xy})^T \\
	&= R_{\mathrm{tom}}(-\Delta \vec{\Omega})^{xy} \cdot (F'')^{xy} \cdot (R_{\mathrm{tom}}(-\Delta \vec{\Omega})^{xy})^T \\
	&= R^T \cdot (F''')^{\ell t} \cdot R
\end{align*} 
\end{widetext}
All three equations yielded same results. Similar verification also holds for other electromagnetic field tensors involved. \par 
Our second approach for verification is based on Ungar et al. \cite{ungar2, ungar3, ungar9} in which we apply direct boost $(\vec{\beta} + \delta \vec{\beta})$ and successive boosts $\vec{\beta}$ and $\Delta \vec{\beta}$ to a position 4-vector. We can check if the results are consistent and share the same overall features as the electromagnetic field tensor. To see that we start with a general position 4-vector in the lab frame and for simplicity, we ignore the time component in the position 4-vector:
\begin{equation} \label{eq:43}
	(r)^{xy} = \begin{pmatrix}
		0 \\
		x \\
		y \\
		z \\
	\end{pmatrix}
\end{equation}
Transforming it in the $\ell t$- frame using the rotation matrix $R$ \eqref{eq:18}, we get:
\begin{align*}
	(r)^{\ell t} &= R \cdot r^{xy} \\
		&=
		\begin{pmatrix}
		0 \\[5pt]
		\frac{x \beta _x+y \beta _y+z \beta _z}{\lambda_1} \\[5pt]
		\frac{y \beta _x-x \beta _y}{\eta_1} \\[5pt]
		\frac{z \beta _x^2-x \beta _z \beta _x+\beta _y \left(z \beta _y-y \beta _z\right)}{\eta_1 \lambda_1} \\ \numberthis \label{eq:44}
		\end{pmatrix}
\end{align*} \par 
We can calculate the expression of $(r)^{\ell t}$ transformed by the first successive boost $\vec{\beta}^{\ell t}$ using Eq. \eqref{eq:21} in the same way we calculated the electromagnetic field tensor $F^{\mu \nu}$:
\begin{align*}
	(r')^{\ell t} &= A(\vec{\beta})^{\ell t} \cdot (r)^{\ell t} \\
		&=
		\begin{pmatrix}
		-\gamma  \left(x \beta _x+y \beta _y+z \beta _z\right) \\[5pt]
		\frac{\gamma  \left(x \beta _x+y \beta _y+z \beta _z\right)}{\lambda_1} \\[5pt]
		\frac{y \beta _x-x \beta _y}{\eta_1} \\[5pt]
		\frac{z \beta _x^2-x \beta _z \beta _x+\beta _y \left(z \beta _y-y \beta _z\right)}{\eta_1 \lambda_1} \\ \numberthis \label{eq:45}
		\end{pmatrix}
\end{align*}
which is nothing but the standard Lorentz transformation of coordinates. Similarly, for the direct boost $(\vec{\beta}+\delta \vec{\beta})^{\ell t}$ and successive boosts $\vec{\beta}^{\ell t}$ and $\Delta \vec{\beta}^{\ell t}$, after letting $\beta_z = \delta \beta_z = 0$ for simplicity, we have: \\
\begin{widetext}
\begin{align*}
(r'')^{\ell t}&= A(\vec{\beta}+\delta \vec{\beta})^{\ell t} \cdot (r)^{\ell t} \\
&=
\begin{pmatrix}
-\frac{\gamma  \left(x \beta _x^3+\left(x \delta \beta _x \gamma ^2+y \beta _y+y \delta \beta _y\right) \beta _x^2+\beta _y \left(x \beta _y+\left(\gamma ^2-1\right) \left(y \delta \beta _x+x \delta \beta _y\right)\right) \beta _x+\beta _y^2 \left(y \delta \beta _y \gamma ^2+y \beta _y+x \delta \beta _x\right)\right)}{\eta_1^2} \\[5pt]
\frac{(\gamma -1) \left(y \beta _x-x \beta _y\right) \left(\beta _x \delta \beta _y-\beta _y \delta \beta _x\right)+\left(x \beta _x+y \beta _y\right) \eta_1^2 \left(\left(\beta _x \delta \beta _x+\beta _y \delta \beta _y\right) \gamma ^3+\gamma \right)}{\eta_1^3} \\[5pt]
\frac{y \beta _x^3-x \left(\beta _y-(\gamma -1) \delta \beta _y\right) \beta _x^2+\beta _y \left(y \beta _y-(\gamma -1) \left(x \delta \beta _x-y \delta \beta _y\right)\right) \beta _x-\beta _y^2 \left(x \beta _y+y (\gamma -1) \delta \beta _x\right)}{\eta_1^3} \\[5pt]
z \\ 
\end{pmatrix} 
\end{align*}

\begin{align*}
	(r''')^{\ell t} &= A(\Delta \vec{\beta})^{\ell t} \cdot A(\vec{\beta})^{\ell t} \cdot (r)^{\ell t} \\
		&= A(\Delta \vec{\beta})^{\ell t}\cdot (r')^{\ell t} \\
		&= \begin{pmatrix}
		\gamma  \left(-\frac{\left(x \beta _x+y \beta _y\right) \left(\beta _x \delta \beta _x+\beta _y \delta \beta _y\right) \gamma ^2}{\eta_1^2}-x \beta _x-y \beta _y-\frac{\left(y \beta _x-x \beta _y\right)\lambda_6}{\eta_1^2}\right) \\[5pt]
		\frac{\gamma  \left(x \beta _x+y \beta _y\right) \left( \beta _x \delta \beta _x \gamma ^2+ \beta _y \delta \beta _y \gamma ^2+ 1 \right)}{\eta_1} \\[5pt]
		\frac{x \gamma ^2 \delta \beta _y \beta _x^2+\left(\beta _y \left(y \delta \beta _y-x \delta \beta _x\right) \gamma ^2+y\right) \beta _x-\beta _y \left(y \beta _y \delta \beta _x \gamma ^2+x\right)}{\eta_1} \\[5pt]
		z \\ \numberthis \label{eq:46}
		\end{pmatrix}
\end{align*}
\end{widetext} \par
One way to check if Eqs. \eqref{eq:45} and \eqref{eq:46} are correct is to show that the invariant interval $ds^2$ \cite{jackson3}:
\begin{align*} 
	ds^2 &= g_{\mu \nu} dx^{\mu} dx^{\nu} \\
		&= (dx^0)^2 - (dx^1)^2 - (dx^2)^2 - (dx^3)^2 \numberthis \label{eq:47}
\end{align*}
remains the same, where $g_{\mu \nu}$ is the metric tensor:
\begin{equation*}
	g_{\mu \nu} = \begin{pmatrix}
	1 & 0 & 0 & 0 \\
	0 & -1 & 0 & 0 \\
	0 & 0 & -1 & 0 \\
	0 & 0 & 0 & -1 \\
	\end{pmatrix}
\end{equation*}
In our case we are concerned with the invariance of 
\begin{equation} \label{eq:48}
	s^2 = x_0^2 - x_1^2 - x_2^2 -x_3^2
\end{equation}
To see if that is the case, we can apply Eq. \eqref{eq:48} to $(r')^{\ell t}$, $(r'')^{\ell t}$ and $(r''')^{\ell t}$ calculated above. The invariant
\begin{equation*}
	s^2 = - x^2 - y^2 - z^2
\end{equation*}
indeed stays the same for each case. This makes sense since we ignored the time component. \par 
Similar results can be obtained for the position 4-vector $r$ in the lab $xy$ -frame and it can be easily proved that the invariant does not change. We also compare our approach with Ungar's in the Appendix (Section C).

\section{Conclusion}
The work presented in this paper is another confirmation of the fact that two successive boosts are not equal to a single direct boost. In the case of the electromagnetic field, just applying the usual electromagnetic field transformation equations will not result in the correct form of electromagnetic fields in the case of non-collinear boosts (accelerating frames) as Thomas rotation must be included. \par  
Apart from the validations made in the previous section we will see if the electromagnetic field tensors in the direct boosted frame $\vec{\beta}+\delta \vec{\beta}$ and the successively boosted frames $\vec{\beta}$ and $\Delta \vec{\beta}$ are consistent with the Thomas rotation. To see that we can take the difference between the corresponding elements of $F''$ and $F'''$ in both the longitudinal-transverse $\ell t$ and lab $xy$-frames. \par 
After taking the difference of the electromagnetic field tensors $F''$ and $F'''$ we found that 
\begin{equation} \label{eq:49}
	(F'')_{ij} - (F''')_{ij} \propto (\gamma -1)
\end{equation}
for both $\ell t$- and $xy$-frames. This makes sense because both $F''$ and $F'''$ just differ by Thomas rotation. Although taking the difference of $F''$ and $F'''$ is not very significant physically it does show what we expected. \par 
To our knowledge, this is the first time that someone has calculated the expressions of the electromagnetic fields in the frames corresponding to general three-dimensional non-collinear boosts.\par 
One application of this work concerns the calculation of shifts in the Larmor frequency of highly relativistic particles moving through non-uniform magnetic and electric fields.  Such a formalism was developed for the motion of non-relativistic particles \cite{golub1, golub2, golub3}; however, this formalism is not directly applicable to relativistic particles because the formalism assumes the electromagnetic fields are known in the particle rest frame.  For a highly relativistic particle undergoing acceleration (e.g., relativistic charged particles stored by electromagnetic fields within a circular storage ring), one can then apply the formalism developed here in this paper to determine the electromagnetic fields in an appropriate reference frame, where any residual motion of the particle is then non-relativistic, and then proceed to calculate the frequency shifts per the formalism of \cite{golub1,golub2,golub3}. \\
\subsection*{Acknowledgments}
This material is based upon work supported by the U.S. Department of Energy, Office of Science, Office of Nuclear Physics, under Award Number DE-SC0014622.
\section{Appendix}
We begin by recalling that the electromagnetic field tensor $F^{\mu \nu}$ is an anti-symmetric matrix: \\
\begin{equation} 
F^{\mu \nu} = \begin{pmatrix}
0 & -E_x & -E_y & -E_z \\
E_x & 0 & -B_z & B_y \\
E_y & B_z & 0 & -B_x \\
E_z & -B_y & B_x & 0 
\end{pmatrix}
\end{equation}
Defining some dummy variables to make equations look cleaner: \\
\begin{align*}
	\lambda_1 &= \sqrt{\beta _x^2+\beta _y^2+\beta _z^2} \\
	\lambda_2 &= \beta_{x} \delta \beta_{x} + \beta_{y} \delta \beta_{y} + \beta_{z} \delta \beta_{z} \\
	\lambda_3 &= \beta _y^2 \delta \beta _x -\beta _x \beta _y \delta \beta _y +\beta _z \left(\beta _z \delta \beta _x-\beta _x \delta \beta _z\right) \\
	\lambda_4 &= \beta _x^2 \delta \beta _y -\beta _x \beta _y \delta \beta _x+\beta _z \left(\beta _z \delta \beta _y-\beta _y \delta \beta _z\right) \\
	\lambda_5 &= \beta _x^2 \delta \beta _z -\beta _x \beta _z \delta \beta _x+\beta _y \left(\beta _y \delta \beta _z-\beta _z \delta \beta _y\right) \\ 
	\lambda_6 &= \beta_{x} \delta \beta_{y} - \beta_{y} \delta \beta_{x} \\
	\lambda_7 &= \beta_{y} \delta \beta_{z} - \beta_{z} \delta \beta_{y} \\
	\lambda_8 &= \beta_{z} \delta \beta_{x} - \beta_{x} \delta \beta_{z} \\	
	\eta_1 &= \sqrt{\beta _x^2+\beta _y^2} \\
	\eta_2 &= \sqrt{\beta _y^2+\beta _z^2} \\
	\eta_3 &= \sqrt{\beta _x^2+\beta _z^2} 
\end{align*} \\
The detailed expressions of the electromagnetic fields (to the first order in $\delta \beta$) can be written as: \\
\begin{widetext}
\subsection{ Longitudinal-Transverse \boldmath{$\ell t$}-Frame}
\subsubsection{\textbf{Direct Boosted Frame \boldmath{$(\vec{\beta}+\delta \vec{\beta})$}}}
\begin{equation*}
\begin{split}
\bm{(F'')^{10}} & = \frac{1}{\lambda_1}\left[\beta _x E_x+\beta _y E_y+\beta _z E_z  -\lambda_3 E_x \gamma ^2-\lambda_4 E_y \gamma ^2-\lambda_5 E_z \gamma ^2+B_z \lambda_6 \gamma + B_y \lambda_8 \gamma +B_x \lambda_7 \gamma +\frac{\gamma (\gamma -1) \eta_2^2 \delta \beta _x E_x}{\lambda_1^2} \right.\\
& \left. +\frac{\gamma (\gamma -1) \eta_3^2 \delta \beta _y E_y}{\lambda_1^2}  +\frac{\gamma (\gamma -1) \eta_1^2 \delta \beta _z E_z}{\lambda_1^2} -\frac{\gamma (\gamma -1) \beta _x \beta _y \delta \beta _y E_x }{\lambda_1^2}-\frac{\gamma (\gamma -1) \beta _x \beta _z \delta \beta _z E_x}{\lambda_1^2}-\frac{\gamma (\gamma -1) \beta _x \beta _y \delta \beta _x E_y}{\lambda_1^2} \right. \\
& \left. -\frac{\gamma (\gamma -1) \beta _y \beta _z \delta \beta _z E_y}{\lambda_1^2}-\frac{\gamma (\gamma -1) \beta _x \beta _z \delta \beta _x E_z}{\lambda_1^2}-\frac{\gamma (\gamma -1) \beta _y \beta _z \delta \beta _y E_z}{\lambda_1^2} \right]
\end{split}
\end{equation*}

\begin{equation*}
\begin{split}
\bm{(F'')^{20}} &= -\gamma  B_z \eta_1 + \frac{\gamma  B_x \beta _x \beta _z}{\eta_1}+\frac{\gamma  B_y \beta _y \beta _z}{\eta_1} -\frac{\gamma  E_x \beta _y}{\eta_1} +\frac{\gamma  \beta _x E_y}{\eta_1} -\frac{\gamma ^3 B_z \beta _x \eta_1 \delta \beta _x}{\lambda_1^2} +\frac{\gamma ^2 \delta \beta _x \beta _y \beta _z E_z}{\eta_1} -\frac{\gamma ^3 B_z \beta _y \eta_1 \delta \beta _y}{\lambda_1^2}\\ 
& +\frac{\gamma  \left(\gamma ^2-1\right) B_x \beta _x^2 \beta _z \delta \beta _x}{\eta_1\lambda_1^2}+\frac{\gamma \left(\gamma ^2-1\right) B_y \beta _x \delta \beta _x \beta _y \beta _z}{\eta_1 \lambda_1^2}-\frac{\gamma  B_z \beta _x \delta \beta _x \beta _z^2}{\eta_1 \lambda_1^2} -\frac{\gamma  B_z \beta _y \delta \beta _y \beta _z^2}{\eta_1 \lambda_1^2}+\frac{\gamma  B_x \beta _x \eta_1 \delta \beta _z}{\lambda_1^2}\\ 
& + \frac{\gamma  \left(\gamma ^2-1\right) B_y \beta _y^2 \delta \beta _y \beta _z}{\eta_1 \lambda_1^2}+\frac{\gamma  \left(\gamma ^2-1\right) B_x \beta _x \beta _y \delta \beta _y \beta _z}{\eta_1 \lambda_1^2} -\frac{\gamma \left(\gamma ^2-1\right) B_z \beta _z \eta_1 \delta \beta _z}{\lambda_1^2} -\frac{(\gamma -1) \gamma ^2 \beta _x \delta \beta _x E_x \beta _y}{\eta_1}\\ 
& + \frac{\gamma ^3 B_x \beta _x \beta _z^2 \delta \beta _z}{\eta_1 \lambda_1^2}+\frac{\gamma ^3 B_y \beta _y \beta _z^2 \delta \beta _z}{\eta_1 \lambda_1^2}+\frac{\gamma  B_y \beta _y \eta_1 \delta \beta _z}{\lambda_1^2} -\frac{\gamma ^3 E_x \beta _y^2 \delta \beta _y}{\eta_1}-\frac{\gamma ^2 \beta _x^2 E_x \delta \beta _y}{\eta_1} -\frac{\gamma ^3 E_x \beta _y \beta _z \delta \beta _z}{\eta_1}+\frac{\gamma ^3 \beta _x^2 \delta \beta _x E_y}{\eta_1}\\
& +\frac{(\gamma -1) \gamma  \beta _x^2 E_x \delta \beta _y}{\eta_1 \lambda_1^2}-\frac{(\gamma -1) \gamma  \beta _x \delta \beta _x E_x \beta _y}{\eta_1 \lambda_1^2} +\frac{(\gamma -1) \gamma ^2 \beta _x \beta _y \delta \beta _y E_y}{\eta_1}-\frac{(\gamma -1) \gamma  \delta \beta _x \beta _y^2 E_y}{\eta_1 \lambda_1^2}  +\frac{\gamma ^2 \delta \beta _x \beta _y^2 E_y}{\eta_1} \\ 
&+\frac{\gamma ^3 \beta _x E_y \beta _z \delta \beta _z}{\eta_1} -\frac{\gamma ^2 \beta _x \delta \beta _y \beta _z E_z}{\eta_1}+\frac{(\gamma -1) \gamma  \beta _x \delta \beta _y \beta _z E_z}{\eta_1 \lambda_1^2}-\frac{(\gamma -1) \gamma  \delta \beta _x \beta _y \beta _z E_z}{\eta_1 \lambda_1^2}+\frac{(\gamma -1) \gamma  \beta _x \beta _y \delta \beta _y E_y}{\eta_1\lambda_1^2} 
\end{split}
\end{equation*}

\begin{equation*}
\begin{split}
\bm{(F'')^{30}} &= \frac{\gamma  B_y \beta _x\lambda_1}{\eta_1}-\frac{\gamma  B_x \beta _y \lambda_1}{\eta_1} -\frac{\gamma  \beta _x E_x \beta _z}{\eta_1 \lambda_1} -\frac{\gamma  \beta _y E_y \beta _z}{\eta_1 \lambda_1} +\frac{\gamma  E_z \eta_1}{\lambda_1} +\frac{\gamma ^3 B_y \beta _x^2 \delta \beta _x}{\eta_1 \lambda_1} +\frac{\gamma  B_y \delta \beta _x \beta _y^2}{\eta_1 \lambda_1} -\frac{\gamma  B_x \beta _x^2 \delta \beta _y}{\eta_1 \lambda_1} \qquad \quad \\ 
&  -\frac{\gamma ^3 B_x \beta _y^2 \delta \beta _y}{\eta_1 \lambda_1}-\frac{\gamma  B_z \beta _z \lambda_6}{\eta_1 \lambda_1} +\frac{\gamma  \left(\gamma ^2-1\right) B_y \beta _x \beta _y \delta \beta _y}{\eta_1 \lambda_1}-\frac{\gamma  \left(\gamma ^2-1\right) B_x \beta _x \delta \beta _x \beta _y}{\eta_1 \lambda_1} -\frac{(\gamma -1) \gamma  \beta _x^2 \delta \beta _x E_x \beta _z}{\eta_1 \lambda_1^3}  \\ 
& +\frac{\gamma ^3 B_y \beta _x \beta _z \delta \beta _z}{\eta_1 \lambda_1}-\frac{\gamma ^3 B_x \beta _y \beta _z \delta \beta _z}{\eta_1 \lambda_1} -\frac{\gamma ^2 \beta _x E_x \delta \beta _z \eta_1}{\lambda_1} -\frac{(\gamma -1) \gamma ^2 \beta _x^2 \delta \beta _x E_x \beta _z}{\eta_1 \lambda_1} -\frac{(\gamma -1) \gamma ^2 \beta _x E_x \beta _y \delta \beta _y \beta _z}{\eta_1 \lambda_1}\\ 
& +\frac{(\gamma -1) \gamma  \beta _x E_x \delta \beta _z \eta_1}{\lambda_1^3}-\frac{(\gamma -1) \gamma  \beta _x E_x \beta _y \delta \beta _y \beta _z}{\eta_1 \lambda_1^3} -\frac{(\gamma -1) \gamma ^2 \beta _x \delta \beta _x \beta _y E_y \beta _z}{\eta_1 \lambda_1}-\frac{(\gamma -1) \gamma ^2 \beta _y^2 \delta \beta _y E_y \beta _z}{\eta_1 \lambda_1}\\ 
& -\frac{\gamma ^3 \beta _x E_x \beta _z^2 \delta \beta _z}{\eta_1 \lambda_1} -\frac{\gamma ^2 \beta _y E_y \delta \beta _z \eta_1}{\lambda_1} +\frac{(\gamma -1) \gamma  \beta _y E_y \lambda_5}{\eta_1 \lambda_1^3}-\frac{\gamma ^3 \beta _y E_y \beta _z^2 \delta \beta _z}{\eta_1 \lambda_1} +\frac{(\gamma -1) \gamma  \beta _z \delta \beta _z E_z \eta_1}{\lambda_1^3} \\ 
& + \frac{\gamma ^3 \beta _y \delta \beta _y E_z \eta_1}{\lambda_1}+\frac{\gamma ^3 \beta _x \delta \beta _x E_z \eta_1}{\lambda_1}+\frac{\gamma ^2 \beta _x \delta \beta _x \beta _z^2 E_z}{\eta_1 \lambda_1}-\frac{(\gamma -1) \gamma  \beta _y \delta \beta _y \beta _z^2 E_z}{\eta_1 \lambda_1^3} + \frac{(\gamma -1) \gamma ^2 \beta _z \delta \beta _z E_z\eta_1^3}{\lambda_1^3}\\ 
& +\frac{\gamma ^2 \beta _y \delta \beta _y \beta _z^2 E_z}{\eta_1 \lambda_1}+\frac{(\gamma -1) \gamma ^2 \beta _z^3 \delta \beta _z E_z \eta_1}{\lambda_1^3} -\frac{(\gamma -1) \gamma  \beta _x \delta \beta _x \beta _z^2 E_z}{\eta_1 \lambda_1^3}
\end{split}
\end{equation*}

\begin{equation*}
\begin{split}
\bm{(F'')^{32}} = \frac{1}{\lambda_1}\left[B_x \beta _x + B_y \beta _y + B_z \beta _z - \gamma  E_x \lambda_7 - \gamma  E_y \lambda_8 - \gamma  E_z \lambda_6 -\frac{(\gamma -1) B_z \lambda_5}{\lambda_1^2}-\frac{(\gamma -1) B_y \lambda_4}{\lambda_1^2} -\frac{(\gamma -1) B_x \lambda_3}{\lambda_1^2} \right] \qquad \quad
\end{split}
\end{equation*}

\begin{equation*}
\begin{split}
\bm{(F'')^{13}} &= -\frac{\gamma  B_x \beta _y}{\eta_1} +\frac{\gamma  B_y \beta _x}{\eta_1} -\frac{\gamma  \beta _x E_x \beta _z}{\eta_1}-\frac{\gamma  \beta _y E_y \beta _z}{\eta_1}+\gamma  E_z \eta_1 -\frac{\gamma ^3 B_x \beta _x \delta \beta _x \beta _y}{\eta_1} -\frac{\gamma ^3 B_x \beta _y^2 \delta \beta _y}{\eta_1} -\frac{\gamma ^3 B_x \beta _y \beta _z \delta \beta _z}{\eta_1}\\ 
& +\frac{\gamma ^3 B_y \beta _x \beta _y \delta \beta _y}{\eta_1}+\frac{\gamma ^3 B_y \beta _x^2 \delta \beta _x}{\eta_1} +\frac{\gamma ^3 B_y \beta _x \beta _z \delta \beta _z}{\eta_1} -\frac{\gamma ^3 \beta _x E_x \beta _z^2 \delta \beta _z}{\eta_1 \lambda_1^2}+\frac{(\gamma -1) B_x \beta _x \delta \beta _x \beta _y}{\eta_1 \lambda_1^2}-\frac{(\gamma -1) B_x \beta _x^2 \delta \beta _y}{\eta_1 \lambda_1^2} \\ 
& +\frac{(\gamma -1) B_y \delta \beta _x \beta _y^2}{\eta_1 \lambda_1^2} +\frac{(\gamma -1) B_z \delta \beta _x \beta _y \beta _z}{\eta_1 \lambda_1^2}-\frac{(\gamma -1) B_y \beta _x \beta _y \delta \beta _y}{\eta_1 \lambda_1^2} -\frac{(\gamma -1) B_z \beta _x \delta \beta _y \beta _z}{\eta_1 \lambda_1^2} -\frac{\left(\gamma ^2-1\right) \gamma  \beta _x^2 \delta \beta _x E_x \beta _z}{\eta_1 \lambda_1^2} \\ 
& -\frac{\left(\gamma ^2-1\right) \gamma  \beta _x E_x \beta _y \delta \beta _y \beta _z}{\eta_1 \lambda_1^2}  -\frac{\gamma  \beta _x E_x \delta \beta _z \eta_1}{\lambda_1^2} -\frac{\gamma  \left(\gamma ^2-1\right) \beta _y^2 \delta \beta _y E_y \beta _z}{\eta_1 \lambda_1^2}-\frac{\gamma  \left(\gamma ^2-1\right) \beta _x \delta \beta _x \beta _y E_y \beta _z}{\eta_1 \lambda_1^2}-\frac{\gamma  \beta _y E_y \delta \beta _z \eta_1}{\lambda_1^2} \\ 
& -\frac{\gamma ^3 \beta _y E_y \beta _z^2 \delta \beta _z}{\eta_1 \lambda_1^2} + \frac{\gamma ^3 \beta _y \delta \beta _y E_z \eta_1}{\lambda_1^2}+\frac{\gamma ^3 \beta _x \delta \beta _x E_z \eta_1}{\lambda_1^2}+\frac{\gamma  \beta _x \delta \beta _x \beta _z^2 E_z}{\eta_1 \lambda_1^2} + \frac{\gamma  \left(\gamma ^2-1\right) \beta _z \delta \beta _z E_z \eta_1}{\lambda_1^2}+\frac{\gamma  \beta _y \delta \beta _y \beta _z^2 E_z}{\eta_1 \lambda_1^2} 
\end{split}
\end{equation*}

\begin{equation*}
\begin{split}
\bm{(F'')^{12}} &= \frac{\gamma  B_z \eta_1}{\lambda_1}-\frac{\gamma  B_x \beta _x \beta _z}{\eta_1 \lambda_1}-\frac{\gamma  B_y \beta _y \beta _z}{\eta_1 \lambda_1} +\frac{\gamma  E_x \beta _y \lambda_1}{\eta_1} -\frac{\gamma  \beta _x E_y \lambda_1}{\eta_1}+\frac{(\gamma -1) B_x \beta _x^2 \delta \beta _x \beta _z}{\eta_1 \lambda_1^3} -\frac{\gamma ^3 B_x \beta _x^2 \delta \beta _x \beta _z}{\eta_1 \lambda_1} -\frac{\gamma  \delta \beta _x \beta _y^2 E_y}{\eta_1 \lambda_1}\\ 
& -\frac{\gamma ^3 B_y \beta _x \delta \beta _x \beta _y \beta _z}{\eta_1 \lambda_1}+\frac{\gamma ^3 B_z \beta _x \delta \beta _x \eta_1}{\lambda_1} + \frac{\gamma ^3 B_z \beta _y \delta \beta _y \eta_1}{\lambda_1}-\frac{\gamma ^3 B_x \beta _x \beta _y \delta \beta _y \beta _z}{\eta_1 \lambda_1}-\frac{\gamma ^3 B_y \beta _y^2 \delta \beta _y \beta _z}{\eta_1 \lambda_1} -\frac{\gamma ^3 B_x \beta _x \beta _z^2 \delta \beta _z}{\eta_1 \lambda_1}\\ 
& +\frac{(\gamma -1) B_z \beta _x \delta \beta _x \beta _z^2}{\eta_1 \lambda_1^3}+\frac{(\gamma -1) B_y \beta _x \delta \beta _x \beta _y \beta _z}{\eta_1 \lambda_1^3}+\frac{(\gamma -1) B_x \beta _x \beta _y \delta \beta _y \beta _z}{\eta_1 \lambda_1^3} +\frac{(\gamma -1) B_y \beta _y^2 \delta \beta _y \beta _z}{\eta_1 \lambda_1^3}+\frac{(\gamma -1) B_z \beta _y \delta \beta _y \beta _z^2}{\eta_1 \lambda_1^3} \\ 
& -\frac{(\gamma -1) B_x \beta _x \delta \beta _z \eta_1}{\lambda_1^3}-\frac{(\gamma -1) B_y \beta _y \delta \beta _z \eta_1}{\lambda_1^3}-\frac{(\gamma -1) B_z \beta _z \delta \beta _z \eta_1}{\lambda_1^3} +\frac{\gamma  \left(\gamma ^2-1\right) \beta _x \delta \beta _x E_x \beta _y}{\eta_1 \lambda_1} -\frac{\left(\gamma ^2-1\right) \gamma  \beta _x \beta _y \delta \beta _y E_y}{\eta_1 \lambda_1}\\ 
& -\frac{\gamma ^3 B_y \beta _y \beta _z^2 \delta \beta _z}{\eta_1 \lambda_1}+\frac{\gamma ^3 B_z \beta _z \delta \beta _z \eta_1}{\lambda_1} + \frac{\gamma ^3 E_x \beta _y^2 \delta \beta _y}{\eta_1 \lambda_1}+\frac{\gamma ^3 E_x \beta _y \beta _z \delta \beta _z}{\eta_1 \lambda_1}+\frac{\gamma  \beta _x^2 E_x \delta \beta _y}{\eta_1 \lambda_1} -\frac{\gamma ^3 \beta _x^2 \delta \beta _x E_y}{\eta_1 \lambda_1}  -\frac{\gamma ^3 \beta _x E_y \beta _z \delta \beta _z}{\eta_1 \lambda_1}\\ 
&-\frac{\gamma  \delta \beta _x \beta _y \beta _z E_z}{\eta_1 \lambda_1} +\frac{\gamma  \beta _x \delta \beta _y \beta _z E_z}{\eta_1 \lambda_1}
\end{split}
\end{equation*}

\subsubsection{\textbf{Successively Boosted Frame \boldmath{$\vec{\beta}$ and $\Delta \vec{\beta}$}}}
\begin{equation*}
\begin{split}
\bm{(F''')^{10}} &= \frac{\beta _x E_x}{\lambda_1} +\frac{\beta _y E_y}{\lambda_1} +\frac{\beta _z E_z}{\lambda_1} +\frac{\gamma ^2 B_z\lambda_6}{\lambda_1}+\frac{\gamma ^2 B_y \lambda_8}{\lambda_1}+\frac{\gamma ^2 B_x \lambda_7}{\lambda_1} +\frac{\gamma ^2 \beta _x E_x \beta _y \delta \beta _y}{\lambda_1}+\frac{\gamma ^2 \beta _x E_x \beta _z \delta \beta _z}{\lambda_1}-\frac{\gamma ^2 \delta \beta _x E_x \eta_2^2}{\lambda_1} \qquad \\ 
& + \frac{\gamma ^2 \beta _y E_y \beta _z \delta \beta _z}{\lambda_1}-\frac{\gamma ^2 \delta \beta _y E_y \eta_3^2}{\lambda_1}+\frac{\gamma ^2 \beta _x \delta \beta _x \beta _y E_y}{\lambda_1} + \frac{\gamma ^2 \beta _y \delta \beta _y \beta _z E_z}{\lambda_1} +\frac{\gamma ^2 \beta _x \delta \beta _x \beta _z E_z}{\lambda_1}-\frac{\gamma ^2 \delta \beta _z E_z \eta_1^2}{\lambda_1}
\end{split}
\end{equation*}

\begin{equation*}
\begin{split}
\bm{(F''')^{20}} &= \frac{2 \gamma  B_y \beta _y \beta _z}{\eta_1}-\gamma  B_z \eta_1 -\frac{\gamma  E_x \beta _y}{\eta_1} +\frac{\gamma  \beta _x E_y}{\eta_1} -\frac{\gamma ^3 B_z \eta_1\lambda_2}{\lambda_1^2}+\frac{\gamma  B_z \beta _z \lambda_5}{\eta_1 \lambda_1^2} -\frac{\gamma ^3 E_x \beta _y\lambda_2}{\eta_1}+ \frac{\gamma ^3 \beta _x E_y\lambda_2}{\eta_1} +\frac{2 \gamma  B_y \beta _y \delta \beta _z \eta_1}{\lambda_1^2}  \quad \quad \\ 
& + \frac{2 \gamma  \left(\gamma ^2-1\right) B_y \beta _y^2 \delta \beta _y \beta _z}{\eta_1 \lambda_1^2}+\frac{2 \gamma  \left(\gamma ^2-1\right) B_y \beta _x \delta \beta _x \beta _y \beta _z}{\eta_1 \lambda_1^2} + \frac{2 \gamma ^3 B_y \beta _y \beta _z^2 \delta \beta _z}{\eta_1 \lambda_1^2}
\end{split}
\end{equation*}

\begin{equation*}
\begin{split}
\bm{(F''')^{30}} &= \frac{\gamma  B_y \beta _x \lambda_1}{\eta_1}-\frac{\gamma  B_x \beta _y \lambda_1}{\eta_1} -\frac{\gamma  \beta _x E_x \beta _z}{\eta_1 \lambda_1} -\frac{\gamma  \beta _y E_y \beta _z}{\eta_1 \lambda_1} +\frac{\gamma  E_z \eta_1}{\lambda_1} -\frac{\gamma  B_x \beta _x \lambda_6}{\eta_1 \lambda_1}-\frac{\gamma  B_y \beta _y \lambda_6}{\eta_1 \lambda_1}-\frac{\gamma  B_z \beta _z \lambda_6}{\eta_1 \lambda_1} +\frac{\gamma ^3 B_y \beta _x\lambda_2}{\eta_1 \lambda_1} \qquad \\ 
& -\frac{\gamma ^3 B_x \beta _y\lambda_2}{\eta_1 \lambda_1} -\frac{\gamma ^3 \beta _x E_x \beta _z\lambda_2}{\eta_1 \lambda_1} -\frac{\gamma ^3 \beta _y E_y \beta _z\lambda_2}{\eta_1 \lambda_1} +\frac{\gamma ^3 E_z \eta_1\lambda_2}{\lambda_1} 
\end{split}
\end{equation*}

\begin{equation*}
\begin{split}
\bm{(F''')^{32}} &= \frac{1}{\lambda_1} \left[B_x \beta _x + B_y \beta _y+B_z \beta _z -\gamma ^2 E_x \lambda_7 - \gamma ^2 E_y \lambda_8 - \gamma ^2 E_z \lambda_6 + \gamma ^2 B_z \beta _x \delta \beta _x \beta _z + \gamma ^2 B_x \beta _x \beta _y \delta \beta _y + \gamma ^2 B_z \beta _y \delta \beta _y \beta _z \right. \qquad \\ 
& \left. - \gamma ^2 B_x \delta \beta _x\eta_2^2 -\gamma ^2 B_y \delta \beta _y \eta_3^2 - \gamma ^2 B_z \delta \beta _z \eta_1^2 + \gamma ^2 B_y \beta _y \beta _z \delta \beta _z + \gamma ^2 B_x \beta _x \beta _z \delta \beta _z + \gamma ^2 B_y \beta _x \delta \beta _x \beta _y \right]
\end{split}
\end{equation*}

\begin{equation*}
\begin{split}
\bm{(F''')^{13}} &= -\frac{\gamma  B_x \beta _y}{\eta_1} +\frac{\gamma  B_y \beta _x}{\eta_1} -\frac{\gamma  \beta _x E_x \beta _z}{\eta_1}-\frac{\gamma  \beta _y E_y \beta _z}{\eta_1}+\gamma  E_z \eta_1 -\frac{\gamma ^3 B_x \beta _y\lambda_2}{\eta_1}+\frac{\gamma ^3 B_y \beta _x\lambda_2}{\eta_1} -\frac{\gamma ^3 \beta _x E_x \beta _z\lambda_2}{\eta_1 \lambda_1^2}-\frac{\gamma ^3 \beta _y E_y \beta _z\lambda_2}{\eta_1 \lambda_1^2} \quad \\ 
& -\frac{\gamma  \beta _x E_x \lambda_5}{\eta_1 \lambda_1^2} - \frac{\gamma  \beta _y E_y \lambda_5}{\eta_1 \lambda_1^2} +\frac{\gamma ^3 E_z \eta_1\lambda_2}{\lambda_1^2} - \frac{\gamma  \beta _z E_z \lambda_5}{\eta_1 \lambda_1^2}
\end{split}
\end{equation*}

\begin{equation*}
\begin{split}
\bm{(F''')^{21}} &= \frac{\gamma  B_z \eta_1}{\lambda_1}-\frac{\gamma  B_x \beta _x \beta _z}{\eta_1 \lambda_1}-\frac{\gamma  B_y \beta _y \beta _z}{\eta_1 \lambda_1} +\frac{\gamma  E_x \beta _y \lambda_1}{\eta_1} - \frac{\gamma  \beta _x E_y \lambda_1}{\eta_1} -\frac{\gamma ^3 B_x \beta _x \beta _z\lambda_2}{\eta_1 \lambda_1}-\frac{\gamma ^3 B_y \beta _y \beta _z\lambda_2}{\eta_1 \lambda_1} +\frac{\gamma ^3 B_z \eta_1\lambda_2}{\lambda_1} +\frac{\gamma ^3 E_x \beta _y\lambda_2}{\eta_1 \lambda_1}  \\ 
& + \frac{\gamma  \beta _x E_x \lambda_6}{\eta_1 \lambda_1} - \frac{\gamma ^3 \beta _x E_y\lambda_2}{\eta_1 \lambda_1} + \frac{\gamma  \beta _y E_y \lambda_6}{\eta_1 \lambda_1} + \frac{\gamma  \beta _z E_z \lambda_6}{\eta_1 \lambda_1} 
\end{split}
\end{equation*}

\subsection{Laboratory \boldmath{$xy$}-Frame}
\subsubsection{\textbf{Direct Boosted Frame \boldmath{$(\vec{\beta}+\delta \vec{\beta})$}}}
	\begin{equation*}
	\begin{split}
	\bm{(F'')^{10}} &= \gamma  B_z \beta _y-\gamma  B_y \beta _z -\gamma ^2 \beta _x^2 E_x +\frac{1}{\lambda_1^2}\left[\gamma  E_x \left(\gamma  \beta_x^2 + \beta_y^2 + \beta_z^2\right)-(\gamma -1) \beta _x \beta _y E_y - (\gamma -1) \beta _x \beta _z E_z -\gamma ^3 B_y \beta _z\lambda_2 + \gamma ^3 B_z \beta _y\lambda_2 \right. \\ 
	& \left. + \gamma  B_z \lambda_4 + \gamma ^3 \beta _x \delta \beta _x E_x \eta_2^2 + \gamma ^3 E_x \beta _y \delta \beta _y \eta_2^2 + \gamma ^3 E_x \beta _z \delta \beta _z \eta_2^2 - \gamma ^3 \beta _x \beta _y \delta \beta _y \beta _z E_z - \gamma ^3 \beta _x \beta _z^2 \delta \beta _z E_z  - \gamma ^3 \beta _x^2 \delta \beta _x \beta _y E_y \right. \\ 
	& \left. -\gamma  B_y \lambda_5 - \gamma ^3 \beta _x \beta _y^2 \delta \beta _y E_y - \gamma ^3 \beta _x \beta _y E_y \beta _z \delta \beta _z - \gamma ^3 \beta _x^2 \delta \beta _x \beta _z E_z +\frac{2 (\gamma -1) \beta _x^2 E_x \beta _y \delta \beta _y}{\lambda_1^2}+\frac{2 (\gamma -1) \beta _x^2 E_x \beta _z \delta \beta _z}{\lambda_1^2}  \right. \\ 
	& \left. -\frac{(\gamma -1) \delta \beta _x \beta _y E_y \left(-\beta_x^2 + \beta_y^2 + \beta_z^2\right)}{\lambda_1^2}  +\frac{2 (\gamma -1) \beta _x \beta _y E_y \beta _z \delta \beta _z}{\lambda_1^2}-\frac{(\gamma -1) \beta _x \delta \beta _y E_y \left(\beta _x^2-\beta _y^2+\beta _z^2\right)}{\lambda_1^2} +\frac{2 (\gamma -1) \beta _x \beta _y \delta \beta _y \beta _z E_z}{\lambda_1^2} \right. \\ 
	& \left. -\frac{(\gamma -1) \delta \beta _x \beta _z E_z \left(-\beta_x^2 + \beta_y^2 + \beta_z^2\right)}{\lambda_1^2}  -\frac{(\gamma -1) \beta _x \delta \beta _z E_z \left(\beta _x^2+\beta _y^2-\beta _z^2\right)}{\lambda_1^2} -\frac{2 (\gamma -1) \beta _x \delta \beta _x E_x \eta_2^2}{\lambda_1^2}\right]
	\end{split}
	\end{equation*}
	
	\begin{equation*}
	\begin{split}
	\bm{(F'')^{20}} &= \gamma  B_x \beta _z-\gamma  B_z \beta _x -\gamma ^2 \beta _x E_x \beta _y -\gamma ^2 \beta _y^2 E_y -\gamma ^2 \beta _y \beta _z E_z + \frac{1}{\lambda_1^2}\left[ \gamma ^2 \beta _y^2 E_y + \gamma  E_y \eta_3^2 + \gamma ^3 \beta _x \delta \beta _x E_y \eta_3^2 + \gamma ^3 \beta _y \delta \beta _y E_y \eta_3^2 \right. \\ 
	& \left. +(\gamma -1) \gamma  \beta _x E_x \beta _y +(\gamma -1) \gamma  \beta _y \beta _z E_z - \gamma ^3 B_z \beta _x^2 \delta \beta _x + \left(\gamma ^2-1\right) \gamma  B_x \beta _x \delta \beta _x \beta _z - \left(\gamma ^2-1\right) \gamma  B_z \beta _x \beta _y \delta \beta _y - \gamma  B_z \delta \beta _x \eta_2^2 \right. \\ 
	& \left. + \gamma ^3 B_x \beta _z^2 \delta \beta _z - \left(\gamma ^2-1\right) \gamma  B_z \beta _x \beta _z \delta \beta _z + \gamma  \left(\gamma ^2-1\right) B_x \beta _y \delta \beta _y \beta _z + \gamma  B_x \delta \beta _z \eta_1^2 - \gamma ^3 \beta _x E_x \beta _y \beta _z \delta \beta _z + \gamma ^3 E_y \beta _z \delta \beta _z \eta_3^2 \right. \\ 
	& \left. -\gamma ^3 \beta _x^2 \delta \beta _x E_x \beta _y - \gamma ^3 \beta _x E_x \beta _y^2 \delta \beta _y - \gamma ^3 \beta _x \delta \beta _x \beta _y \beta _z E_z - \gamma ^3 \beta _y^2 \delta \beta _y \beta _z E_z - \gamma ^3 \beta _y \beta _z^2 \delta \beta _z E_z -\frac{(\gamma -1) \beta _y \delta \beta _z E_z \left(\beta _x^2+\beta _y^2-\beta _z^2\right)}{\lambda_1^2} \right. \\ 
	& \left. -\frac{(\gamma -1) \delta \beta _x E_x \beta _y \left(-\beta_x^2 + \beta_y^2 + \beta_z^2\right)}{\lambda_1^2}-\frac{(\gamma -1) \beta _x E_x \delta \beta _y \left(\beta _x^2-\beta _y^2+\beta _z^2\right)}{\lambda_1^2} +\frac{2 (\gamma -1) \beta _x \delta \beta _x \beta _y^2 E_y}{\lambda_1^2} -\frac{2 (\gamma -1) \beta _y \delta \beta _y E_y \eta_3^2}{\lambda_1^2} \right. \\ 
	& \left. +\frac{2 (\gamma -1) \beta _y^2 E_y \beta _z \delta \beta _z}{\lambda_1^2} +\frac{2 (\gamma -1) \beta _x \delta \beta _x \beta _y \beta _z E_z}{\lambda_1^2} -\frac{(\gamma -1) \delta \beta _y \beta _z E_z \left(\beta _x^2-\beta _y^2+\beta _z^2\right)}{\lambda_1^2} +\frac{2 (\gamma -1) \beta _x E_x \beta _y \beta _z \delta \beta _z}{\lambda_1^2} \right] 
	\end{split}
	\end{equation*}
	
	\begin{equation*}
	\begin{split}
	\bm{(F'')^{30}} &= \gamma  B_y \beta _x-\gamma  B_x \beta _y-\gamma ^2 \beta _z^2 E_z + \frac{1}{\lambda_1^2}\left[\gamma ^3 B_y \beta _x^2 \delta \beta _x - (\gamma -1) \beta _x E_x \beta _z - (\gamma -1) \beta _y E_y \beta _z + \gamma  E_z \left(\beta _x^2+\beta _y^2+\gamma  \beta _z^2\right) \right. \\ 
	& \left. + \gamma ^3 B_y \beta _x \beta _y \delta \beta _y - \gamma ^3 B_x \beta _y^2 \delta \beta _y - \left(\gamma ^2-1\right) \gamma  B_x \beta _x \delta \beta _x \beta _y + \gamma  B_y \delta \beta _x \eta_2^2 - \gamma  B_y \beta _x \beta _y \delta \beta _y + \gamma ^3 B_y \beta _x \beta _z \delta \beta _z \right.  \\
	& \left. - \gamma ^3 B_x \beta _y \beta _z \delta \beta _z + \gamma  B_x \beta _y \beta _z \delta \beta _z - \gamma  B_x \delta \beta _y \eta_3^2 - \gamma  B_y \beta _x \beta _z \delta \beta _z - \gamma ^3 \beta _x^2 \delta \beta _x E_x \beta _z - \gamma ^3 \beta _x E_x \beta _y \delta \beta _y \beta _z - \gamma ^3 \beta _y E_y \beta _z^2 \delta \beta _z \right. \\
	& \left. - \gamma ^3 \beta _x E_x \beta _z^2 \delta \beta _z - \gamma ^3 \beta _x \delta \beta _x \beta _y E_y \beta _z - \gamma ^3 \beta _y^2 \delta \beta _y E_y \beta _z + \gamma ^3 \beta _x \delta \beta _x E_z \eta_1^2 + \gamma ^3 \beta _y \delta \beta _y E_z \eta_1^2 + \gamma ^3 \beta _z \delta \beta _z E_z \eta_1^2 \right. \\ 
	& \left. - \frac{(\gamma -1) \delta \beta _x E_x \beta _z \left(-\beta_x^2 + \beta_y^2 + \beta_z^2\right)}{\lambda_1^2}+\frac{2 (\gamma -1) \beta _x E_x \beta _y \delta \beta _y \beta _z}{\lambda_1^2} -\frac{(\gamma -1) \beta _x E_x \delta \beta _z \left(\beta _x^2+\beta _y^2-\beta _z^2\right)}{\lambda_1^2}+\frac{2 (\gamma -1) \beta _x \delta \beta _x \beta _y E_y \beta _z}{\lambda_1^2} \right. \\ 
	& \left. -\frac{(\gamma -1) \delta \beta _y E_y \beta _z \left(\beta _x^2-\beta _y^2+\beta _z^2\right)}{\lambda_1^2} -\frac{(\gamma -1) \beta _y E_y \delta \beta _z \left(\beta _x^2+\beta _y^2-\beta _z^2\right)}{\lambda_1^2} +\frac{2 (\gamma -1) \beta _x \delta \beta _x \beta _z^2 E_z}{\lambda_1^2} +\frac{2 (\gamma -1) \beta _y \delta \beta _y \beta _z^2 E_z}{\lambda_1^2} \right. \\ 
	& \left.- \frac{2 (\gamma -1) \beta _z \delta \beta _z E_z \eta_1^2}{\lambda_1^2} \right]
	\end{split}
	\end{equation*}
	
	\begin{equation*}
	\begin{split}
	\bm{(F'')^{32}} &= \gamma  E_y \beta _z -\gamma  \beta _y E_z + \frac{1}{\lambda_1^2} \left[B_x \beta _x^2 + \gamma  B_x \eta_2^2 - (\gamma -1) B_y \beta _x \beta _y - (\gamma -1) B_z \beta _x \beta_z - \gamma ^3 B_y \beta _x^2 \delta \beta _x \beta _y - \gamma ^3 B_z \beta _x^2 \delta \beta _x \beta _z \right. \qquad \quad \\ 
	& \left. +\gamma ^3 B_x \beta _x \delta \beta _x \eta_2^2 + \gamma ^3 B_x \beta _y \delta \beta _y \eta_2^2 - \gamma ^3 B_y \beta _x \beta _y^2 \delta \beta _y - \gamma ^3 B_z \beta _x \beta _y \delta \beta _y \beta _z + \gamma ^3 B_x \beta _z \delta \beta _z \eta_2^2 - \gamma ^3 B_z \beta _x \beta _z^2 \delta \beta _z \right. \\ 
	& \left. -\gamma ^3 B_y \beta _x \beta _y \beta _z \delta \beta _z - \gamma  \delta \beta _y E_z \eta_3^2 + \gamma ^3 E_y \beta _z^2 \delta \beta _z + \left(\gamma ^2-1\right) \gamma  \beta _y \delta \beta _y E_y \beta _z + \gamma  \left(\gamma ^2-1\right) \beta _x \delta \beta _x E_y \beta _z + \gamma  E_y \delta \beta _z \eta_1^2 \right. \\ 
	& \left. -\gamma ^3 \beta _y^2 \delta \beta _y E_z - \left(\gamma ^2-1\right) \gamma  \beta _x \delta \beta _x \beta _y E_z - \left(\gamma ^2-1\right) \gamma  \beta _y \beta _z \delta \beta _z E_z - \frac{2 (\gamma -1) B_x \beta _x \delta \beta _x \eta_2^2}{\lambda_1^2} +\frac{2 (\gamma -1) B_x \beta _x^2 \beta _z \delta \beta _z}{\lambda_1^2} \right. \\ 
	& \left. +\frac{2 (\gamma -1) B_y \beta _x \beta _y \beta _z \delta \beta _z}{\lambda_1^2} +\frac{2 (\gamma -1) B_x \beta _x^2 \beta _y \delta \beta _y}{\lambda_1^2}-\frac{(\gamma -1) B_z \delta \beta _x \beta _z \left(-\beta_x^2 + \beta_y^2 + \beta_z^2\right)}{\lambda_1^2} +\frac{2 (\gamma -1) B_z \beta _x \beta _y \delta \beta _y \beta _z}{\lambda_1^2} \right.\\ 
	& \left. -\frac{(\gamma -1) B_y \beta _x \delta \beta _y \left(\beta _x^2-\beta _y^2+\beta _z^2\right)}{\lambda_1^2} -\frac{(\gamma -1) B_y \delta \beta _x \beta _y \left(-\beta_x^2 + \beta_y^2 + \beta_z^2\right)}{\lambda_1^2} - \frac{(\gamma -1) B_z \beta _x \delta \beta _z \left(\beta _x^2+\beta _y^2-\beta _z^2\right)}{\lambda_1^2} \right]
	\end{split}
	\end{equation*}
	
	\begin{equation*}
	\begin{split}
	\bm{(F'')^{13}} &=  -\gamma  E_x \beta _z +\gamma  \beta _x E_z + \frac{1}{\lambda_1^2}\left[B_y \beta _y^2 + \gamma  B_y \eta_3^2 - (\gamma -1) B_z \beta _y \beta _z - (\gamma -1) B_x \beta _x \beta _y - \gamma ^3 B_x \beta _x^2 \delta \beta _x \beta _y - \gamma ^3 B_z \beta _y^2 \delta \beta _y \beta _z \right. \\
	& \left. -\gamma ^3 B_x \beta _x \beta _y^2 \delta \beta _y - \gamma ^3 B_x \beta _x \beta _y \beta _z \delta \beta _z + \gamma ^3 \beta _x \beta _z \delta \beta _z E_z + \gamma ^3 \beta _x^2 \delta \beta _x E_z + \gamma  \delta \beta _x E_z \eta_2^2 - \gamma  \beta _x \beta _y \delta \beta _y E_z - \gamma  \beta _x \beta _z \delta \beta _z E_z \right. \\ 
	& \left. -\gamma ^3 E_x \beta _z^2 \delta \beta _z - \left(\gamma ^2-1\right) \gamma  \beta _x \delta \beta _x E_x \beta _z - \left(\gamma ^2-1\right) \gamma  E_x \beta _y \delta \beta _y \beta _z - \gamma  E_x \delta \beta _z \eta_1^2 + \gamma ^3 \beta _x \beta _y \delta \beta _y E_z - \gamma ^3 B_z \beta _y \beta _z^2 \delta \beta _z \right. \\ 
	& \left. +\gamma ^3 B_y \beta _x \delta \beta _x \eta_3^2 + \gamma ^3 B_y \beta _z \delta \beta _z \eta_3^2 + \gamma ^3 B_y \beta _y \delta \beta _y \eta_3^2 - \gamma ^3 B_z \beta _x \delta \beta _x \beta _y \beta _z + \frac{2 (\gamma -1) B_y \beta _x \delta \beta _x \beta _y^2}{\lambda_1^2}-\frac{2 (\gamma -1) B_y \beta _y \delta \beta _y \eta_3^2}{\lambda_1^2} \right. \\ 
	& \left. +\frac{2 (\gamma -1) B_y \beta _y^2 \beta _z \delta \beta _z}{\lambda_1^2} +\frac{2 (\gamma -1) B_z \beta _x \delta \beta _x \beta _y \beta _z}{\lambda_1^2} -\frac{(\gamma -1) B_z \delta \beta _y \beta _z \left(\beta _x^2-\beta _y^2+\beta _z^2\right)}{\lambda_1^2}-\frac{(\gamma -1) B_z \beta _y \delta \beta _z \left(\beta _x^2+\beta _y^2-\beta _z^2\right)}{\lambda_1^2} \right. \\ 
	&  \left. +\frac{2 (\gamma -1) B_x \beta _x \beta _y \beta _z \delta \beta _z}{\lambda_1^2}-\frac{(\gamma -1) B_x \delta \beta _x \beta _y \left(-\beta_x^2 + \beta_y^2 + \beta_z^2\right)}{\lambda_1^2} -\frac{(\gamma -1) B_x \beta _x \delta \beta _y \left(\beta _x^2-\beta _y^2+\beta _z^2\right)}{\lambda_1^2} \right]
	\end{split}
	\end{equation*}
	
	\begin{equation*}
	\begin{split}
	\bm{(F'')^{21}} &= \gamma  E_x \beta _y -\gamma  \beta _x E_y + \frac{1}{\lambda_1^2}\left[B_z \beta _z^2 + \gamma  B_z \eta_1^2 - (\gamma -1) B_x \beta _x \beta _z - (\gamma -1) B_y \beta _y \beta _z + \gamma ^3 B_z \beta _x \delta \beta _x \eta_1^2 - \gamma ^3 B_x \beta _x^2 \delta \beta _x \beta _z \right. \qquad \quad \\ 
	& \left. -\gamma ^3 \beta _x^2 \delta \beta _x E_y - \gamma ^3 \beta _x \beta _y \delta \beta _y E_y - \gamma ^3 \beta _x E_y \beta _z \delta \beta _z + \gamma  \beta _x \beta _y \delta \beta _y E_y + \gamma  \beta _x E_y \beta _z \delta \beta_z - \gamma  \delta \beta _x E_y \eta_2^2 + \gamma  E_x \delta \beta _y \eta_3^2 \right. \\
	& \left. + \left(\gamma ^2-1\right) \gamma  E_x \beta _y \beta _z \delta \beta _z + \gamma  \left(\gamma ^2-1\right) \beta _x \delta \beta _x E_x \beta _y - \gamma ^3 B_y \beta _x \delta \beta _x \beta _y \beta _z + \gamma ^3 B_z \beta _y \delta \beta _y \eta_1^2 - \gamma ^3 B_y \beta _y^2 \delta \beta _y \beta _z \right. \\ 
	& \left. -\gamma ^3 B_x \beta _x \beta _y \delta \beta _y \beta _z + \gamma ^3 B_z \beta _z \delta \beta _z \eta_1^2 - \gamma ^3 B_x \beta _x \beta _z^2 \delta \beta _z - \gamma ^3 B_y \beta _y \beta _z^2 \delta \beta _z + \gamma ^3 E_x \beta _y^2 \delta \beta _y + \frac{2 (\gamma -1) B_z \beta _y \delta \beta _y \beta _z^2}{\lambda_1^2} \right. \\ 
	& \left. +\frac{2 (\gamma -1) B_y \beta _x \delta \beta _x \beta _y \beta _z}{\lambda_1^2}+\frac{2 (\gamma -1) B_z \beta _x \delta \beta _x \beta _z^2}{\lambda_1^2}-\frac{(\gamma -1) B_y \delta \beta _y \beta _z \left(\beta _x^2-\beta _y^2+\beta _z^2\right)}{\lambda_1^2} -\frac{2 (\gamma -1) B_z \beta _z \delta \beta _z \eta_1^2}{\lambda_1^2} \right. \\ 
	& \left. -\frac{(\gamma -1) B_x \beta _x \delta \beta _z \left(\beta _x^2+\beta _y^2-\beta _z^2\right)}{\lambda_1^2}  -\frac{(\gamma -1) B_x \delta \beta _x \beta _z \left(-\beta_x^2 + \beta_y^2 + \beta_z^2\right)}{\lambda_1^2} +\frac{2 (\gamma -1) B_x \beta _x \beta _y \delta \beta _y \beta _z}{\lambda_1^2} \right. \\ 
	& \left. -\frac{(\gamma -1) B_y \beta _y \delta \beta _z \left(\beta _x^2+\beta _y^2-\beta _z^2\right)}{\lambda_1^2} \right]
	\end{split}
	\end{equation*}
	
	\subsubsection{\textbf{Successively Boosted Frame \boldmath{$\vec{\beta}$ and $\Delta \vec{\beta}$}}}
	
	\begin{equation*}
	\begin{split}
	\bm{(F''')^{10}} &= \gamma  B_z \beta _y-\gamma  B_y \beta _z -\gamma ^2 \beta _x^2 E_x -\gamma ^2 \beta _x \beta _z E_z -\gamma ^2 \beta _x \beta _y E_y + \frac{1}{\lambda_1^2}\left[ \gamma ^2 \beta _x^2 E_x + \gamma  E_x \eta_2^2 + \gamma ^2 \beta _x \beta _y E_y - \gamma  \beta _x \beta _y E_y - \gamma ^2 \beta _x E_z \lambda_5 \right. \\
	& \left. + \gamma ^2 \beta _x \beta _z E_z - \gamma  \beta _x \beta _z E_z + \gamma ^2 B_y \beta _x \delta \beta _x \beta _z - \gamma ^2 B_z \beta _x \delta \beta _x \beta _y + \gamma  B_y \beta _y \delta \beta _y \beta _z - \gamma  B_z \beta _y \beta _z \delta \beta _z - (\gamma -1) \gamma  B_x \beta _x \delta \beta _y \beta _z \right. \\ 
	& \left. -\gamma ^3 B_y \beta _z\lambda_2 + \gamma  B_z \delta \beta _y \left(\gamma  \beta _x^2+\beta _z^2\right) - \gamma  B_y \delta \beta _z \left(\gamma  \beta _x^2+\beta _y^2\right) - \gamma ^2 \beta _x E_x \lambda_3 - \gamma ^3 \beta _x \beta _y E_y\lambda_2 + (\gamma -1) \gamma  B_x \beta _x \beta _y \delta \beta _z \right. \\ 
	& \left. -\gamma ^2 \beta _x E_y \lambda_4 - \gamma ^3 \beta _x \beta _z E_z\lambda_2 + \gamma ^3 E_x \eta_2^2\lambda_2 + \gamma ^3 B_z \beta _y\lambda_2 \right]
	\end{split}
	\end{equation*}
	
	\begin{equation*}
	\begin{split}
	\bm{(F''')^{20}} &= \gamma  B_x \beta _z-\gamma  B_z \beta _x -\gamma ^2 \beta _x E_x \beta _y -\gamma ^2 \beta _y^2 E_y -\gamma ^2 \beta _y \beta _z E_z + \frac{1}{\lambda_1^2}\left[ \gamma ^3 B_x \beta _y \delta \beta _y \beta _z(\gamma -1) \gamma  \beta _x E_x \beta _y + (\gamma -1) \gamma  \beta _y \beta _z E_z \right.   \\ 
	& \left. + \gamma ^3 B_x \beta _x \delta \beta _x \beta _z - \gamma ^3 B_z \beta _x^2 \delta \beta _x + \gamma ^2 B_z \beta _x \beta _y \delta \beta _y + (\gamma -1) \gamma  B_y \delta \beta _x \beta _y \beta _z - \gamma  B_x \beta _x \delta \beta _x \beta _z + \gamma ^2 \beta _x \delta \beta _x \beta _y^2 E_y - \gamma ^3 \beta _y^2 \delta \beta _y \beta _z E_z \right. \\ 
	& \left.  + \gamma  E_y \left(\beta _x^2+\gamma  \beta _y^2+\beta _z^2\right)- \gamma ^3 B_z \beta _x \beta _y \delta \beta _y - \gamma ^2 B_x \beta _y \delta \beta _y \beta _z + \gamma  B_x \delta \beta _z \left(\beta _x^2+\gamma  \beta _y^2\right) + \gamma  B_z \beta _x \beta _z \delta \beta _z - (\gamma -1) \gamma  B_y \beta _x \beta _y \delta \beta _z \right. \\ 
	& \left. - \gamma ^3 B_z \beta _x \beta _z \delta \beta _z + \gamma ^3 B_x \beta _z^2 \delta \beta _z - \gamma ^3 \beta _x^2 \delta \beta _x E_x \beta _y - \gamma ^3 \beta _x E_x \beta _y^2 \delta \beta _y + \gamma ^2 \beta _x E_x \beta _y^2 \delta \beta _y - \gamma ^2 \delta \beta _x E_x \beta _y \eta_2^2 + \gamma ^2 \beta _x E_x \beta _y \beta _z \delta \beta _z \right. \\ 
	& \left. - \gamma ^3 \beta _x E_x \beta _y \beta _z \delta \beta _z + \gamma ^3 \beta _x \delta \beta _x E_y \eta_3^2 - \gamma ^2 \beta _y \delta \beta _y E_y \eta_3^2 + \gamma ^3 \beta _y \delta \beta _y E_y \eta_3^2 + \gamma ^2 \beta _y^2 E_y \beta _z \delta \beta _z + \gamma ^3 E_y \beta _z \delta \beta _z \eta_3^2 - \gamma ^2 \beta _y \delta \beta _z E_z\eta_1^2 \right. \\ 
	& \left. - \gamma ^3 \beta _x \delta \beta _x \beta _y \beta _z E_z - \gamma ^3 \beta _y \beta _z^2 \delta \beta _z E_z + \gamma ^2 \beta _y^2 \delta \beta _y \beta _z E_z + \gamma ^2 \beta _x \delta \beta _x \beta _y \beta _z E_z - \gamma  B_z \delta \beta _x \left(\gamma  \beta _y^2+\beta _z^2\right) \right]
	\end{split}
	\end{equation*}
	
	\begin{equation*}
	\begin{split}
	\bm{(F''')^{30}} &= \gamma  B_y \beta _x-\gamma  B_x \beta _y -\gamma ^2 \beta _x E_x \beta _z -\gamma ^2 \beta _y E_y \beta _z -\gamma ^2 \beta _z^2 E_z + \frac{1}{\lambda_1^2}\left[ \gamma ^3 B_y \beta _x^2 \delta \beta _x (\gamma -1) \gamma  \beta _x E_x \beta _z + (\gamma -1) \gamma  \beta _y E_y \beta _z  \right.\\
	& \left. + \gamma  E_z \left(\beta _x^2+\beta _y^2+\gamma  \beta _z^2\right)- \gamma ^3 B_x \beta _x \delta \beta _x \beta _y + \gamma  B_y \delta \beta _x \left(\beta _y^2+\gamma  \beta _z^2\right) + \gamma  B_x \beta _x \delta \beta _x \beta _y - (\gamma -1) \gamma  B_z \delta \beta _x \beta _y \beta _z - \gamma  B_y \beta _x \beta _y \delta \beta _y \right. \\ 
	& \left. + \gamma ^3 B_y \beta _x \beta _y \delta \beta _y - \gamma ^2 B_y \beta _x \beta _z \delta \beta _z + (\gamma -1) \gamma  B_z \beta _x \delta \beta _y \beta _z - \gamma  B_x \delta \beta _y \left(\beta _x^2+\gamma  \beta _z^2\right) - \gamma ^3 B_x \beta _y \beta _z \delta \beta _z + \gamma ^3 B_y \beta _x \beta _z \delta \beta _z \right. \\ 
	& \left. + \gamma ^2 B_x \beta _y \beta _z \delta \beta _z - \gamma ^3 \beta _x^2 \delta \beta _x E_x \beta _z - \gamma ^3 \beta _x E_x \beta _y \delta \beta _y \beta _z + \gamma ^2 \beta _x E_x \beta _y \delta \beta _y \beta _z - \gamma ^2 \delta \beta _x E_x \beta _z \eta_2^2 + \gamma ^2 \beta _x E_x \beta _z^2 \delta \beta _z \right. \\ 
	& \left. - \gamma ^3 \beta _x E_x \beta _z^2 \delta \beta _z - \gamma ^3 \beta _x \delta \beta _x \beta _y E_y \beta _z - 2 \gamma ^3 \beta _y^2 \delta \beta _y E_y \beta _z - 2 \gamma ^2 \delta \beta _y E_y \beta _z \eta_3^2 + \gamma ^2 \beta _x \delta \beta _x \beta _y E_y \beta _z + \gamma ^2 \beta _y E_y \beta _z^2 \delta \beta _z \right. \\ 
	& \left. - \gamma ^3 \beta _y E_y \beta _z^2 \delta \beta _z + \gamma ^3 \beta _y \delta \beta _y E_z \eta_1^2 + \gamma ^3 \beta _z \delta \beta _z E_z \eta_1^2 + \gamma ^3 \beta _x \delta \beta _x E_z \eta_1^2 + \gamma ^2 \beta _x \delta \beta _x \beta _z^2 E_z - \gamma ^2 \beta _z \delta \beta _z E_z \eta_1^2 - \gamma ^3 B_x \beta _y^2 \delta \beta _y \right. \\ 
	&  \left. + \gamma ^2 \beta _y \delta \beta _y \beta _z^2 E_z \right]
	\end{split}
	\end{equation*}
	
	\begin{equation*}
	\begin{split}
	\bm{(F''')^{32}} &= \gamma  E_y \beta _z-\gamma  \beta _y E_z +\gamma ^3 \beta _y \delta \beta _y E_y \beta _z-\gamma ^3 \beta _y^2 E_y \delta \beta _z +\gamma ^3 \delta \beta _y \beta _z^2 E_z -\gamma ^3 \beta _y \beta _z \delta \beta _z E_z + \frac{1}{\lambda_1^2}\left[ - \gamma ^2 B_y \beta _x \lambda_4 + \gamma ^3 B_x \eta_2^2 \lambda_2 \qquad \quad \right. \\ 
	& \left. - (\gamma -1) B_z \beta _x \beta _z + B_x \left(\beta _x^2+\gamma  \left(\beta _y^2+\beta _z^2\right)\right) + (\gamma -1) \gamma  \beta _x E_x \delta \beta _y \beta _z - (\gamma -1) \gamma  \beta _x E_x \beta _y \delta \beta _z - \gamma ^2 B_z \beta _x \lambda_5 - \gamma ^3 B_z \beta _x \beta _z\lambda_2 \right. \\ 
	& \left. + \gamma ^2 E_y \delta \beta _z \left(\beta _x^2+\gamma  \left(\beta _y^2+\beta _z^2\right)\right) - \gamma ^2 \delta \beta _y E_z \left(\beta _x^2+\gamma  \left(\beta _y^2+\beta _z^2\right)\right) (\gamma -1) \gamma ^2 \beta _x \delta \beta _x E_y \beta _z - (\gamma -1) B_y \beta _x \beta _y- \gamma ^3 B_y \beta _x \beta _y\lambda_2   \right. \\ 
	& \left. - \gamma ^2 B_x \beta _x \lambda_3- (\gamma -1) \gamma ^2 \beta _x \delta \beta _x \beta _y E_z \right]  
	\end{split}
	\end{equation*}
	
	\begin{equation*}
	\begin{split}
	\bm{(F''')^{13}} &= \gamma  \beta _x E_z -\gamma  E_x \beta _z - \gamma ^3 \beta _x \delta \beta _x E_x \beta _z +\gamma ^3 \beta _x^2 E_x \delta \beta _z +\gamma ^3 \beta _x \beta _z \delta \beta _z E_z + \frac{1}{\lambda_1^2}\left[\gamma ^3 B_y \eta_3^2 \lambda_2 - \gamma ^2 B_x \beta _y \lambda_3 - \gamma ^3 B_x \beta _x \beta _y\lambda_2 \qquad \quad \right. \\ 
	& \left. + \gamma  \delta \beta _x \beta _z^2 E_z -(\gamma -1) B_x \beta _x \beta _y - (\gamma -1) B_z \beta _y \beta _z  - \gamma ^2 B_y \beta _y \lambda_4 - \gamma ^2 \delta \beta _x \beta _y E_y \beta _z - \gamma ^2 B_z \beta _y \lambda_5 - \gamma  \beta _x \beta _y E_y \delta \beta _z \right. \\ 
	& \left. - \gamma ^3 B_z \beta _y \beta _z\lambda_2 - \gamma ^2 E_x \beta _y^2 \delta \beta _z - (\gamma -1) \gamma ^2 E_x \beta _y \delta \beta _y \beta _z - \gamma ^3 E_x \delta \beta _z \eta_3^2 + \gamma ^2 \beta _x \beta _y E_y \delta \beta _z + \gamma ^3 \beta _x^2 \delta \beta _x E_z + \gamma ^2 \delta \beta _x \beta _y^2 E_z \right. \\ 
	& \left. + \gamma  \delta \beta _x \beta _y E_y \beta _z + (\gamma -1) \gamma ^2 \beta _x \beta _y \delta \beta _y E_z + B_y \left(\gamma  \beta _x^2+\beta _y^2+\gamma  \beta _z^2\right)\right] 
	\end{split}
	\end{equation*}
	
	\begin{equation*}
	\begin{split}
	\bm{(F''')^{21}} &= \gamma ^3 \beta _x \delta \beta _x E_x \beta _y-\gamma ^3 \beta _x^2 E_x \delta \beta _y+\gamma  E_x \beta _y +\gamma ^3 \delta \beta _x \beta _y^2 E_y -\gamma  \beta _x E_y -\gamma ^3 \beta _x \beta _y \delta \beta _y E_y +\gamma ^3 \delta \beta _x \beta _y \beta _z E_z -\gamma ^3 \beta _x \delta \beta _y \beta _z E_z \qquad \\ 
	& + \frac{1}{\lambda_1^2}\left[\gamma ^3 B_z \eta_1^2 \lambda_2 - (\gamma -1) B_x \beta _x \beta _z - (\gamma -1) B_y \beta _y \beta _z + B_z \left(\gamma  \beta _x^2+\gamma  \beta _y^2+\beta _z^2\right) - \gamma ^3 B_x \beta _x \beta _z\lambda_2 - \gamma ^3 B_y \beta _y \beta _z\lambda_2 \right. \\ 
	& \left. + \gamma ^3 E_x \delta \beta _y \eta_1^2 - \gamma ^3 \delta \beta _x E_y \eta_1^2 - \gamma ^2 B_x \beta _z \lambda_3 - \gamma ^2 B_y \beta _z \lambda_4 - \gamma ^2 B_z \beta _z \lambda_5 + (\gamma -1) \gamma ^2 E_x \beta _y \beta _z \delta \beta _z - (\gamma -1) \gamma ^2 \beta _x E_y \beta _z \delta \beta _z \right. \\ 
	& \left. + (\gamma -1) \gamma ^2 \beta _z E_z \lambda_6 + \gamma ^2 E_x \delta \beta _y \beta _z^2 - \gamma ^2 \delta \beta _x E_y \beta _z^2 \right]
	\end{split}
	\end{equation*}
	\end{widetext}

	\subsection{Comparison with the References}
	As mentioned in the Introduction, this work is inspired by Ungar et al. \cite{ungar2,ungar3, ungar4, ungar5, ungar6, ungar7, ungar9} so it is natural to compare our approach with Ungar to see if we share the same overall features. To check that we will work for a 2-dimensional case for simplicity by letting $\beta_y = \beta_z = \delta \beta_z = 0$ in Eq. \eqref{eq:38} which will give \cite{jackson5}:
	\begin{equation} \label{eq:50}
	A(\Delta \vec{\beta}) = \begin{pmatrix}
	1 & -\gamma ^2 \delta \beta _x  & -\gamma \delta \beta _y & 0 \\
	-\gamma ^2 \delta \beta _x  & 1 & 0 & 0\\
	-\gamma \delta \beta _y & 0 & 1 & 0 \\
	0 & 0 & 0 & 1 \\
	\end{pmatrix}
	\end{equation}
	From $A(\Delta \vec{\beta})$, the relativistic velocity composition can be easily extracted:
	\begin{equation} \label{eq:51}
	\Delta \vec{\beta} = \gamma^2 \delta \beta_x \hat{x}+ \gamma \delta \beta_y \hat{y}
	\end{equation}
	Now for the three different inertial frames discussed in the Introduction $\Sigma$, $\Sigma'$ and $\Sigma''$, we have 
	\begin{center}
		Relative velocity of $\Sigma'$ with respect to $\Sigma : \vec{\beta}$  \\
		Relative velocity of $\Sigma''$ with respect to $\Sigma' : \Delta \vec{\beta}$
	\end{center}
	We can now use the relativistic velocity composition rule mentioned in Eq. \eqref{eq:8} to calculate the relative velocity of $\Sigma''$ with respect to $\Sigma$ \cite{ungar2,ungar3, ungar4, ungar5, ungar6, ungar7, ungar9} by defining:
	\begin{align*}
	\vec{u} &= \vec{\beta}  = \beta \hat{x} \\
	\vec{v} &= \Delta \vec{\beta} = \gamma^2 \delta\beta_{x} \hat{x}+ \gamma \delta \beta_{y} \hat{y} \\
	\gamma_u &= \gamma \\
	\gamma_v &\approx 1\\
	\end{align*}
	After plugging in the values and simplifying, we get:
	\begin{align*}
	(\vec{u} \oplus \vec{v})_x &= \frac{\beta + \gamma^2 \delta \beta_x}{1+\gamma^2 \beta \delta \beta_x} \\
	(\vec{u} \oplus \vec{v})_y &= \frac{\delta \beta_y}{1+\gamma^2 \beta \delta \beta_x} \\
	\gamma_{\vec{u} \oplus \vec{v}} &\approx \gamma (1+\gamma^2 \beta \delta \beta_x) \numberthis \label{eq:52}
	\end{align*} \par 
	Constructing the boost matrix from the above equation:
	\begin{widetext}
		\begin{equation*}
		A(\vec{\beta}+\delta \vec{\beta}) = 
		\begin{pmatrix}
		\gamma + \gamma^3 \beta \delta\beta_x  & -(\gamma \beta+\gamma^3 \delta\beta_x)  & -\gamma  \delta\beta _y & 0 \\
		-(\gamma \beta+\gamma^3 \delta\beta_x) & \gamma + \gamma^3 \beta \delta\beta_x & (\gamma -1)\frac{ \delta \beta _y}{\beta} & 0 \\
		-\gamma  \delta\beta _y & (\gamma -1)\frac{\delta \beta _y}{\beta} & 1 & 0 \\
		0 & 0 & 0 & 1\\
		\end{pmatrix}
		\end{equation*}
	\end{widetext}
	We can compare this boost matrix with Eq. \eqref{eq:40} and notice that they are completely identical. Therefore our approach gives the identical final result as discussed in \cite{ungar2,ungar3, ungar4, ungar5, ungar6, ungar7, ungar9}.

\end{document}